\documentclass[11pt,a4paper]{article} 
\usepackage{graphicx}
\usepackage{jcappub}

\newcommand{\simgt}{\lower.5ex\hbox{$\; \buildrel > \over \sim \;$}}
\newcommand{\simlt}{\lower.5ex\hbox{$\; \buildrel < \over \sim \;$}}

\makeatletter
\newsavebox{\@parc@ption}
\def\parcaption#1{%
\sbox{\@parc@ption}{\shortstack[l]{#1}}%
>\setbox\@tempboxa\hbox{\csname fnum@\@captype\endcsname}%
\@tempdima\columnwidth \advance\@tempdima-\wd\@tempboxa
\@tempdimb\@tempdima 
\ifdim\wd\@parc@ption>\@tempdimb \@tempdima\@tempdimb
\else\@tempdima\wd\@parc@ption\fi
\sbox{\@tempboxa}{\parbox[t]{\@tempdima}{#1}}%
\caption{\usebox{\@tempboxa}}}
\makeatother

\title{
{Impacts of satellite galaxies \\on the redshift-space distortions}
}

\author{Chiaki Hikage$^{1}$ and Kazuhiro Yamamoto$^{2,3}$}

\affiliation{
$^{1}$Kobayashi-Maskawa Institute, Nagoya University, Nagoya 464-8602, Japan\\
$^{2}$Department of Physical Sciences, Hiroshima University, Higashi-hiroshima, \\Kagamiyama 1-3-1, 739-8526, Japan\\
$^{3}$Hiroshima Astrophysical Science Center, Hiroshima University,  Higashi-Hiroshima, \\Kagamiyama 1-3-1, 739-8526, Japan
}

\emailAdd{hikage@kmi.nagoya-u.ac.jp, kazuhiro@hiroshima-u.ac.jp}

\abstract{ We study the impacts of the satellite galaxies on the
  redshift-space distortions.  In our multipole power spectrum
  analysis of the luminous red galaxies (LRGs) samples of the Sloan
  digital sky survey (SDSS), we have clearly detected the non-zero
  signature of the hexadecapole and tetrahexadecapole spectrum, which
  almost disappears in the power spectrum with the sample of the
  brightest LRGs only. We thus demonstrate that the satellite LRGs in
  multiple systems make a significant contribution to the multipole
  power spectrum though its fraction is small.  The behavior can be
  understood by a simple halo model, in which the one-halo term,
  describing the Finger of God (FoG) effect from the satellite
  galaxies, makes the dominant contribution to the higher multipole
  spectra.  We demonstrate that the small-scale information of higher
  multipole spectrum is useful for calibrating the satellite FoG
  effect and improves the measurement of the cosmic growth rate
  dramatically. We further demonstrate that the fiber collision in the
  galaxy survey influences the one-halo term and the higher multipole
  spectra, because the number of satellite galaxies in the halo
  occupation distribution (HOD) is changed.  We also discuss about the
  impact of satellite galaxies on future high-redshift surveys
  targeting the H-alpha emitters.  } 
\arxivnumber{1303.3380}

\begin{document}
\maketitle

\flushbottom

\section{Introduction} 
\label{sec:intro}
The luminous red galaxies (LRGs) in the Sloan digital sky survey
(SDSS) demonstrated the usefulness of a large redshift-survey of
galaxies.  Especially, it proved that a precise measurement of the
statistical features in their spatial distribution provides us with
the very useful methodology not only for the cosmology but also for
the fundamental physics.  For example, the baryon acoustic oscillation
signature in the large scale structure is now recognized as a
promising way for exploring the origin of the accelerated expansion of
the universe \cite{Percival2010,Cabre2}.  A stringent constraint on the
neutrino mass is also obtained \cite{Reid2010,Saito2011}.  Furthermore,
the LRG sample showed that a measurement of the redshift-space
distortions gives us a unique chance of testing the theory of gravity
(e.g., \cite{gr,fr}, cf. \cite{Guzzo2008}).

The LRGs in SDSS are massive early-type galaxies, and most part of
them are considered to be residing in the center of halos.  However,
it is clarified that the some fraction of the LRGs consists of
multiple galaxies system. The halo occupation distribution (HOD) of
the LRGs was clarified by Reid and Spergel (\cite{Reid2009a},
cf. \cite{Zheng2009}).  The correspondence between the LRGs and halos
has been investigated and has illuminated the importance of the Finger
of God (FoG) effect \cite{Reid2010,Reid2009b,Hikage2011,Hikage2012},
which is non-linear redshift distortion due to the internal motion of
galaxies within halos \cite{Jackson1972}.  Recent investigations with
N-body simulations have discovered that halos at the redshift 2 could
be the origin of the LRG host halos~\cite{Masaki2012}.  In the present
paper, we investigate the contribution of the satellite LRGs in
multiple systems to the redshift-space distortions.  A related topic
has been investigated in the literature
\cite{Reid2010,Reid2009b,Hikage2011,Hikage2012}, but the previous
works investigated the contribution to the monopole spectrum.  We here
focus our investigation on the redshift-space distortions described by
the higher multipole power spectrum.

The redshift-space distortions are measured in terms of the
anisotropic correlation function or the anisotropic power spectrum,
e.g., \cite{Peacock2001,Zehavi2002,Guzzo2008}.  The anisotropic
correlation function of the SDSS LRG sample has been measured in the
literature, e.g., \cite{Cabre2,Cabre1,Okumura2008}.  The anisotropic
power spectrum $P(k,\mu)$, where $\mu$ denotes the directional cosine
between the line of sight direction and the wave number vector, is the
Fourier transform of the anisotropic correlation function. They are
equivalent to each other.  The multipole power spectrum $P_\ell(k)$ is
defined as the coefficient of the multipole expansion,\footnote{Note
  that the definition of $P_\ell(k)$ is different from the
  conventional definition by the factor $2\ell+1$.}
\footnote{There are also
different works using phase-space distribution function approach to
study the redshift distortion effect on the multipole power spectrum
\cite{SeljakMcDonald2011,Okumura2012}.}
\begin{eqnarray}
P(k,\mu)=\sum_\ell P_\ell(k) {\cal L}_\ell(\mu)(2\ell+1),
\label{defpl1}
\end{eqnarray}
or
\begin{eqnarray}
P_\ell(k)={1\over 2}\int_{-1}^{+1}P(k,\mu){\cal L}_\ell(\mu)d\mu,
\label{defpl2}
\end{eqnarray}
where ${\cal L}_\ell(\mu)$ is the Legendre polynomial, which is normalized as
\begin{eqnarray}
\int_{-1}^{+1}{\cal L}_\ell(\mu){\cal L}_{\ell'}(\mu)d\mu={2\over 2\ell +1}\delta_{\ell\ell'}.
\end{eqnarray}
In refs.~\cite{gr,sato}, the multipole power spectrum of the SDSS LRG
sample was measured.  In the present paper, we demonstrate that the
satellite galaxies make a significant contribution to the higher 
multipole power spectrum, though its fraction is small. 

The primary purpose of the present paper is to understand the
contribution of satellite galaxies to the multipole power spectrum.
To this end, we measure the multipole power spectrum of the SDSS LRG
samples, and compare the results with the predictions of a simple halo
model with the HOD of the SDSS LRG catalog.  Then, we show the
importance of the one halo term in the higher multipole power
spectrum. We demonstrate that the information of multipole power
spectra such as hexadecapole $P_4(k)$ and tetrahexadecapole
$P_6(k)$ are useful for calibrating the satellite properties and
significantly improve the measurement of the growth rate.  We also
investigate the influence of satellite galaxies in a future redshift
survey targeting H-alpha emitters on the multipole power spectrum in a
measurement of the redshift-space distortions, because their
contamination could give rise to a systematic error when comparing
with theoretical models. An assessment of the systematic error is also
the purpose of the present paper.

This paper is organized as follows.  In section \ref{sec:LRG}, we show
the multipole power spectrum of the satellite LRGs and their
contribution to the total LRG sample and the impact on the growth rate
measurement.  In section \ref{sec:an}, we introduce a simple halo
model for a system consisting of central galaxies and satellite
galaxies, then we show that the halo model with the HOD of LRGs
explains the behavior of the LRG multipole spectra.
\ref{sec:constraint}, we demonstrate a constraint from the LRG samples
by comparing the observational results and the theoretical model.  In
section \ref{sec:forecast}, we also discuss about the impact of
satellite galaxies in a future survey targeting H-alpha emitters at
high redshifts.  Section \ref{sec:conclusion} is devoted to summary
and conclusions.  Appendix outlines the derivation of our theoretical
expression for the multipole power spectrum in redshift space in the
halo model.  Throughout the present paper, we adopt the Hubble
constant $H_0=100h$km/s/Mpc with $h=0.7$ unless otherwise stated.

\begin{figure}[b]
\begin{center}
\includegraphics[scale=0.55]{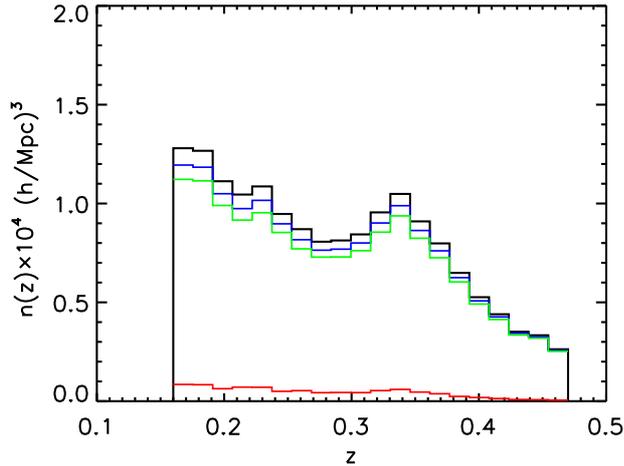}
\caption{Histogram of the number density of the LRG samples.
The black, blue, green and red curves show the number density
of the All, BLRG, Single, and NBLRG, respectively.   
\label{fig:zdist}
}
\end{center}
\end{figure}

\begin{table*}[h]
\begin{center}
\begin{tabular}{ccc}
\hline
\hline
Name of LRG sample & Total Number of LRGs & Number of non-brightest LRGs \\
\hline
All & 96762 & 4716 \\
BLRG & 92046 & 0 \\
Single & 87889  & 0 \\
NBLRG & 0  & 4716 \\
\hline
\end{tabular}
\caption{Properties of LRG samples: ``All'' include all of LRGs in the
  SDSS DR7 LRG Sample in Northern sky; ``BLRG'' includes the brightest LRG in each
  group and the fainter LRGs are excluded; ``Single'' includes LRGs in
  single systems only, and any LRGs in multiple LRG systems are
  excluded; ``NBLRG'' consists of LRGs in multiple LRG systems
  except for the brightest LRGs.
\label{tab:lrg_halo}
}
\end{center}
\end{table*}

\section{Impacts of satellite galaxies on the redshift-space distortions} 
\label{sec:LRG}
In this section, we demonstrate the contribution of the satellite LRGs
to the multipole power spectrum.  We here use the halo sample
described in \cite{Hikage2012} using observed SDSS DR7-Full LRG sample
in Northern sky (publicly available catalog prepared by
\cite{Kazin2010}). The sample consists of 96762 LRGs with the
magnitude $-23.2<M_g<-21.2$ in the redshift range $0.16<z<0.47$ (the
mean redshift is 0.32) covering 1.44(Gpc/h)$^3$ comoving volume. Halo
is identified with the counts-in-cylinders techniques developed by
\cite{Reid2009a}: two galaxies are considered neighbors when the
transverse separation $\Delta r_\perp\le 0.8$Mpc/h and the redshift
difference $\Delta z/(1+z)\le 0.006$ corresponding to the velocity
difference $\delta v_p=1800$km/s.  The total number of halos is 92046.
When the missing galaxies due to fiber collisions are taken into
account, the total number of LRGs become 98991. If all of them are
hosted by the same halos of the observed LRGs, the actual number of
satellite LRGs becomes 6945 (7\%).  Most of halos (95.5\%) occupy
single LRG (hereafter we call them ``single LRG systems'') and the
rest of them contain multiple LRGs (``multiple LRG systems''). The
multiplicity distribution of LRGs in halos is listed in Table 1 of
\cite{Hikage2012}. For the multiple LRG systems, we choose the
brightest LRG (BLRG) in each group as the central LRG and the rest of
them are the non-brightest LRGs (NBLRGs), which we regard as satellite
LRGs. Strictly speaking, BLRGs are not always central LRGs as
suggested by several observations (e.g., \cite{Hikage2012}), and our
satellite sample contains central LRGs to some extent. We have used
different samples described in Table 1 to see the impact of satellite
galaxies on the redshift-space power spectrum.

Figure~\ref{fig:zdist} shows the histogram of the number density of 
the galaxy samples as a function of the redshift $z$. 

\begin{figure*}[t]
\begin{center}
\includegraphics[scale=.45]{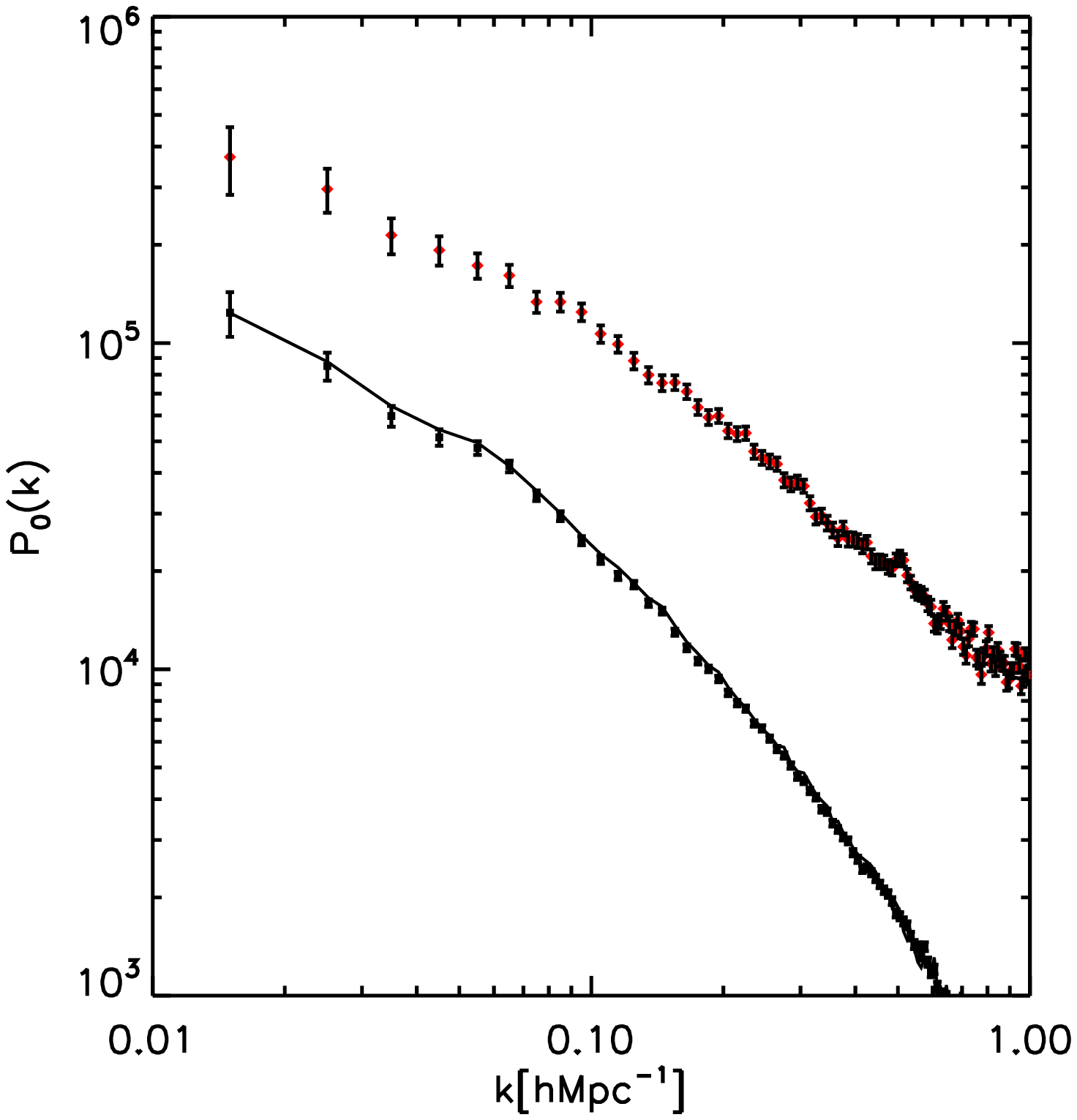}
\includegraphics[scale=.45]{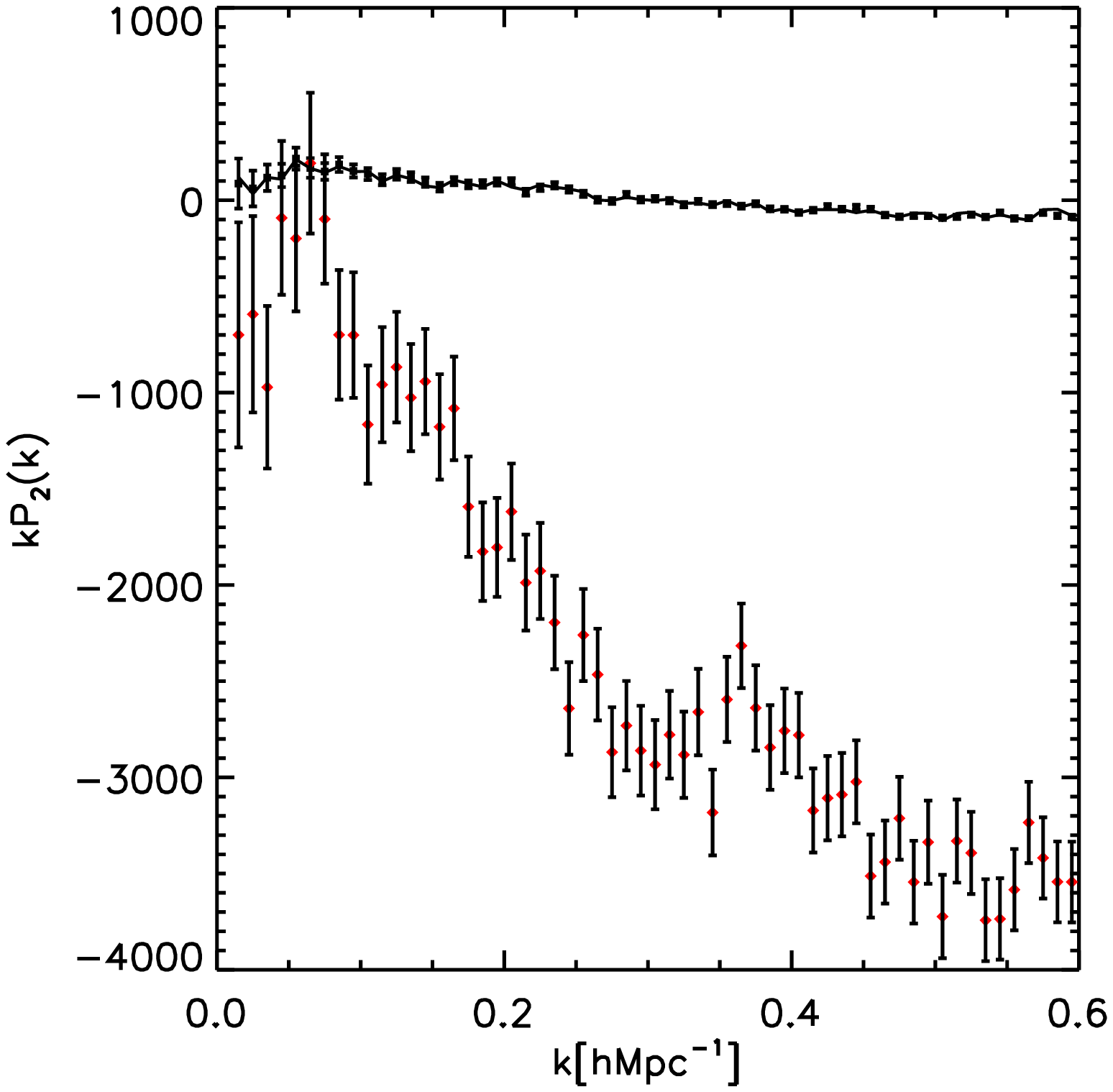}
\includegraphics[scale=.45]{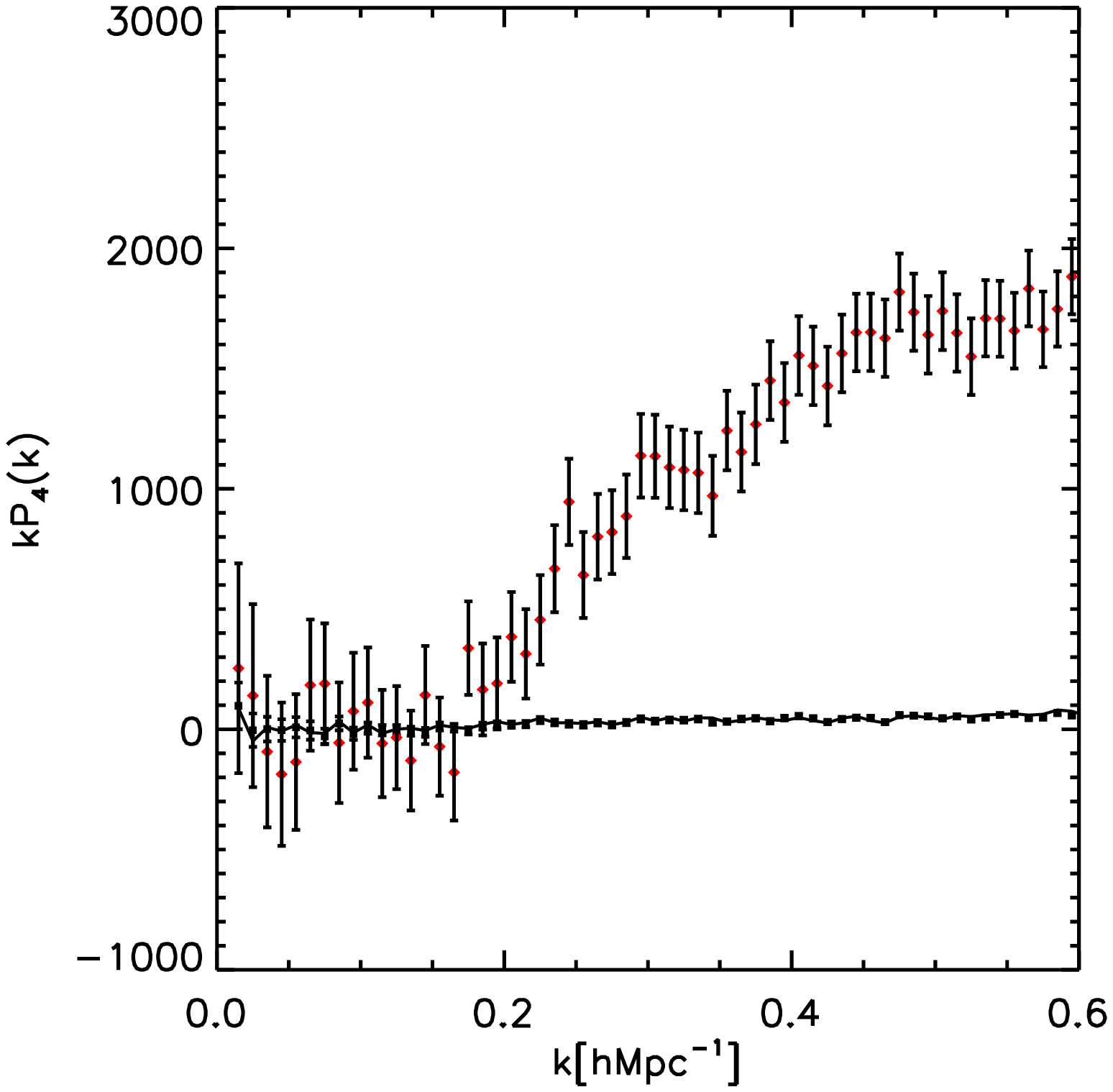}
\includegraphics[scale=.45]{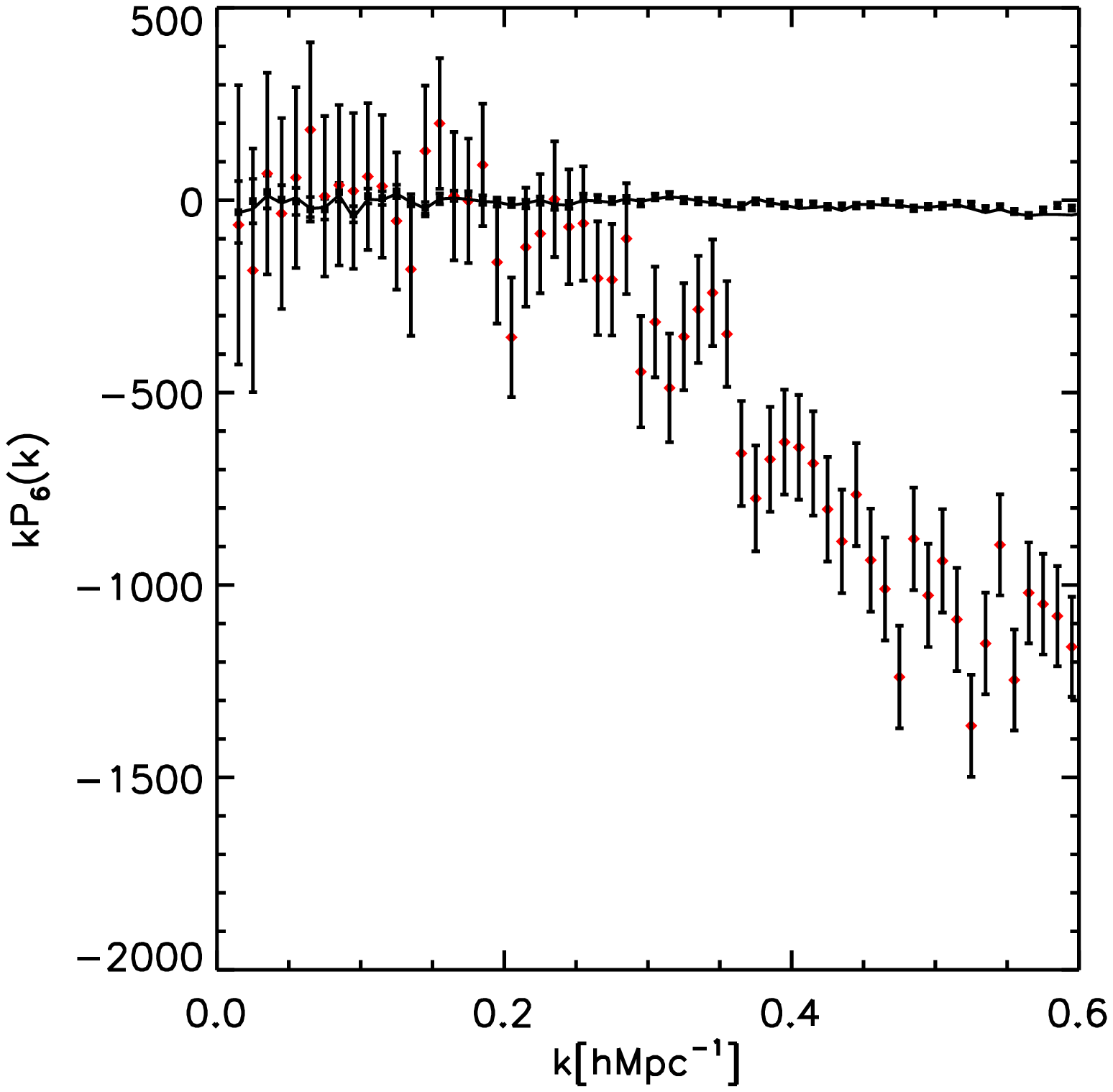}
\caption{Multipole power spectra $P_0(k)$, $P_2(k)$, $P_4(k)$, and $P_6(k)$
for the All LRG sample (black curve) and for the NBLRG (red diamond 
with large error bars). The squares with the small error bars show the 
results with the sample in a previous paper for comparison.
\label{fig:satellite}
}
\end{center}
\end{figure*}

\begin{figure*}[t]
\begin{center}
\includegraphics[scale=.45]{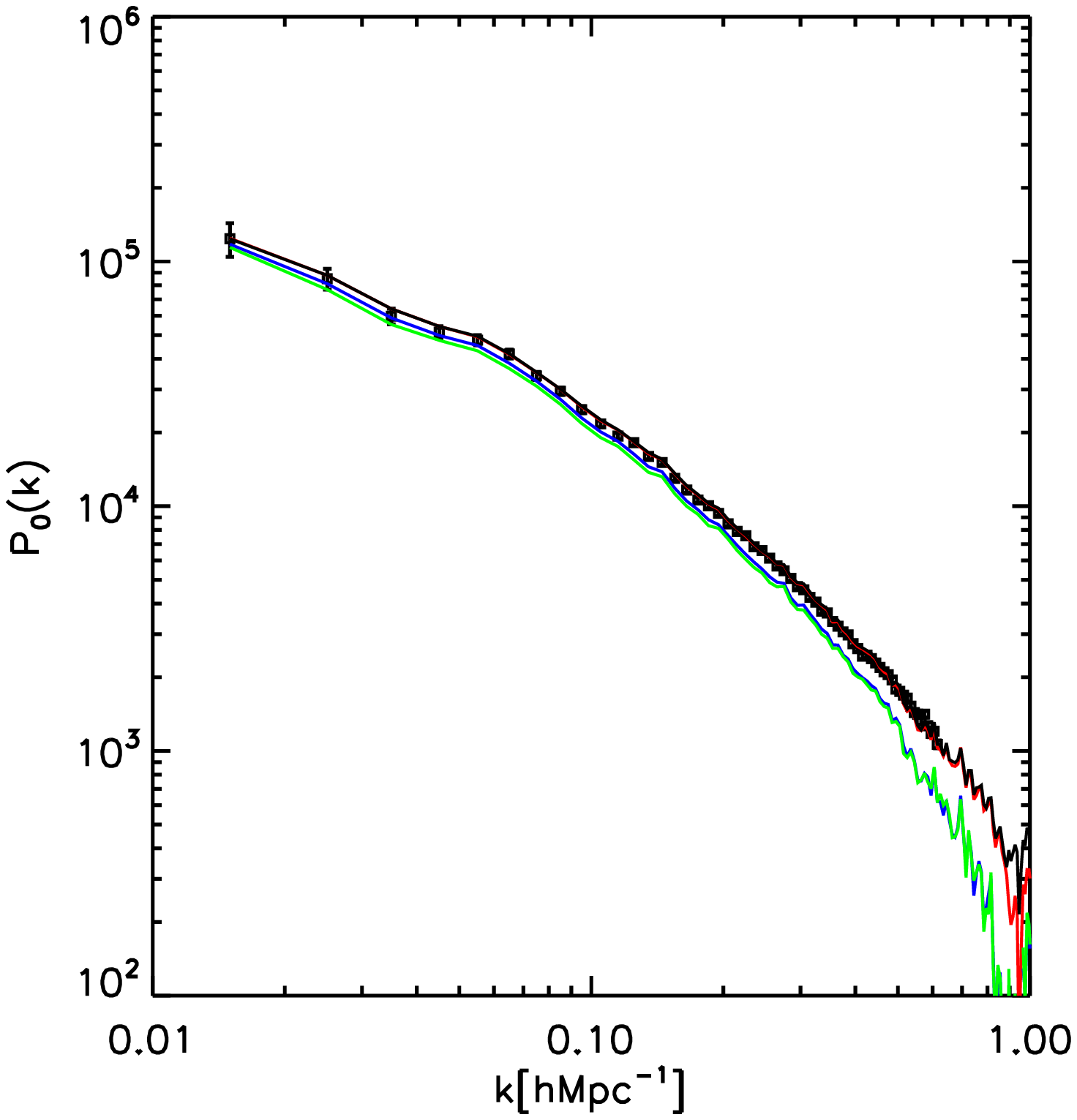}
\includegraphics[scale=.45]{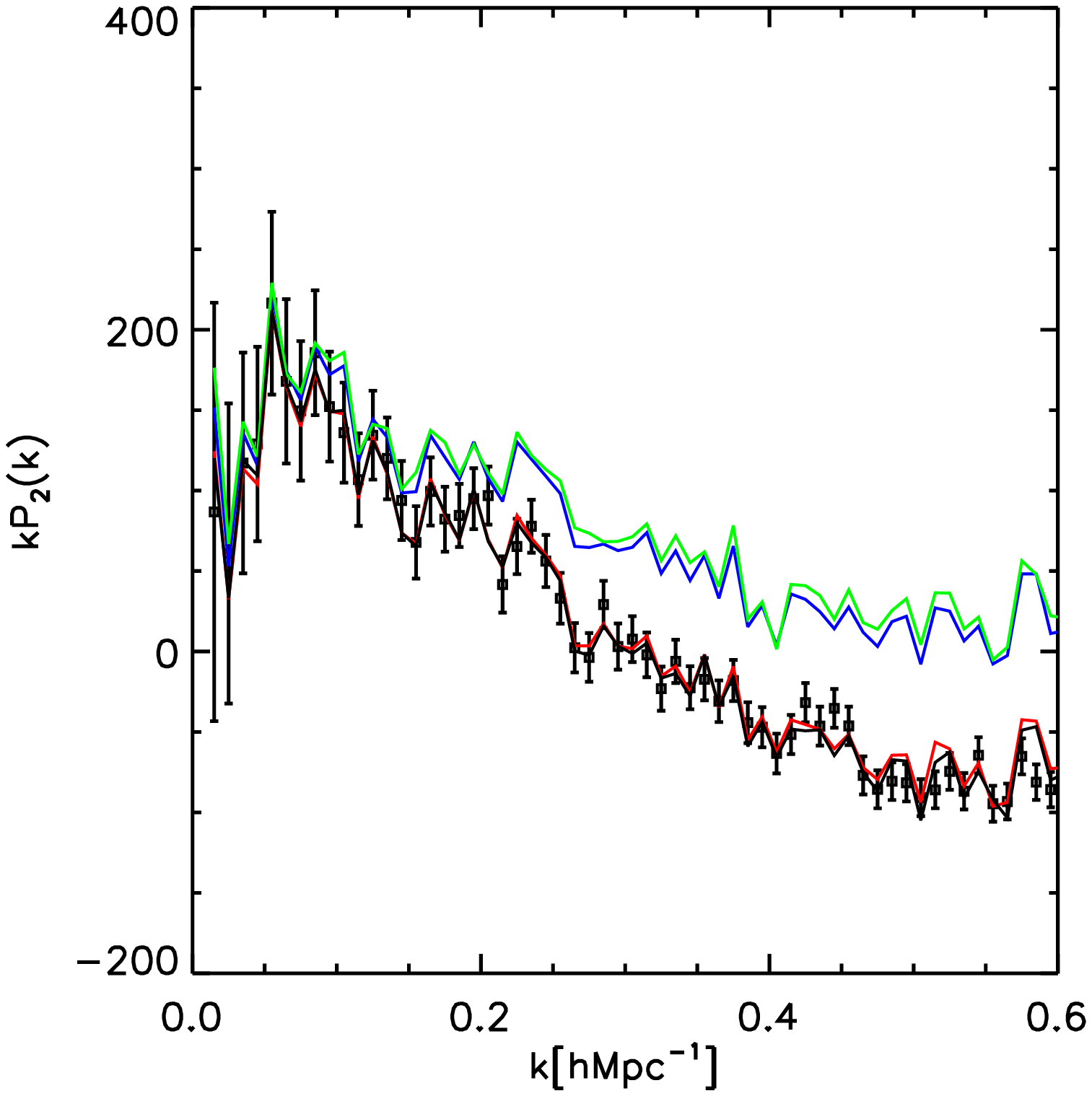}
\includegraphics[scale=.45]{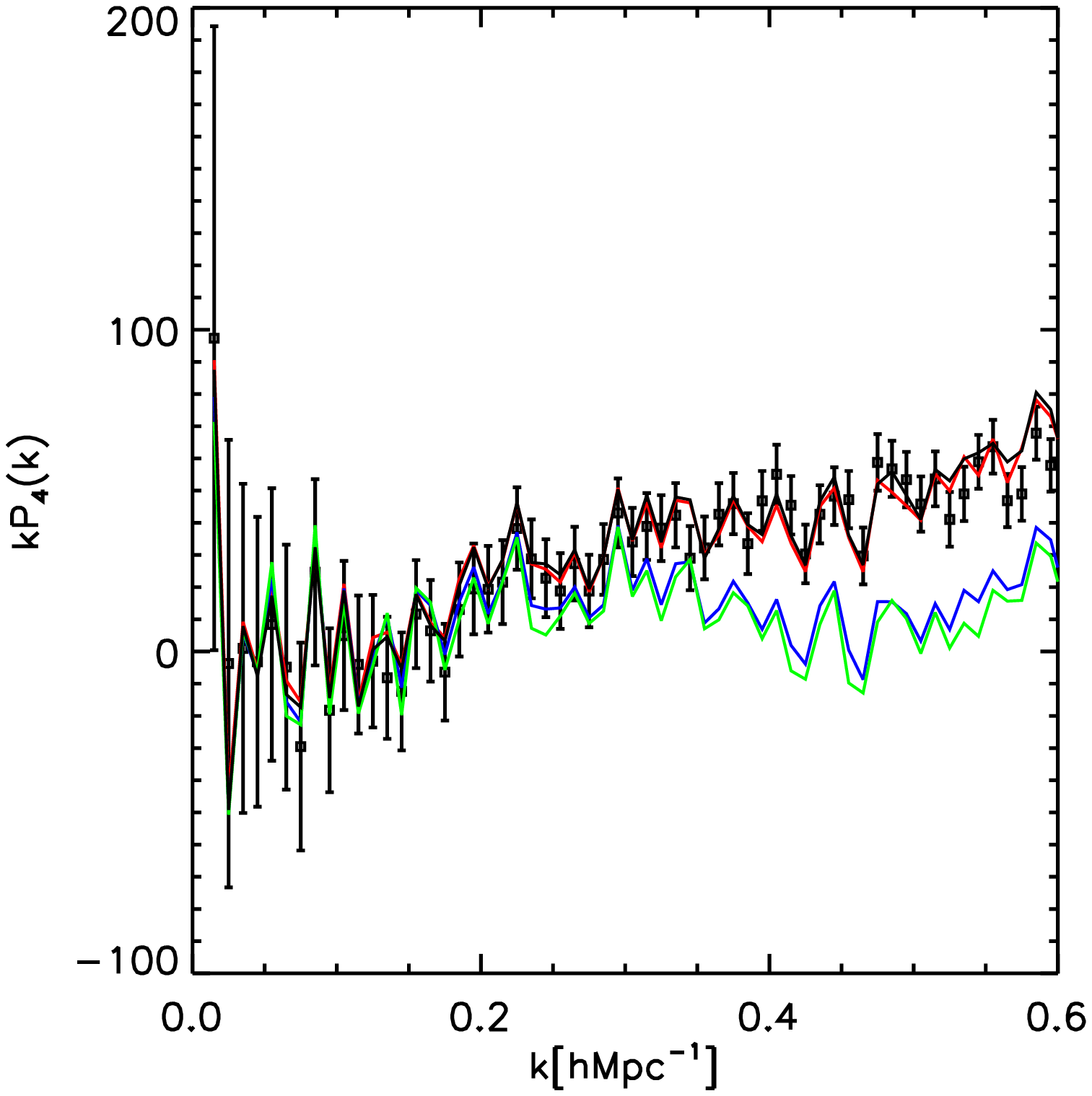}
\includegraphics[scale=.45]{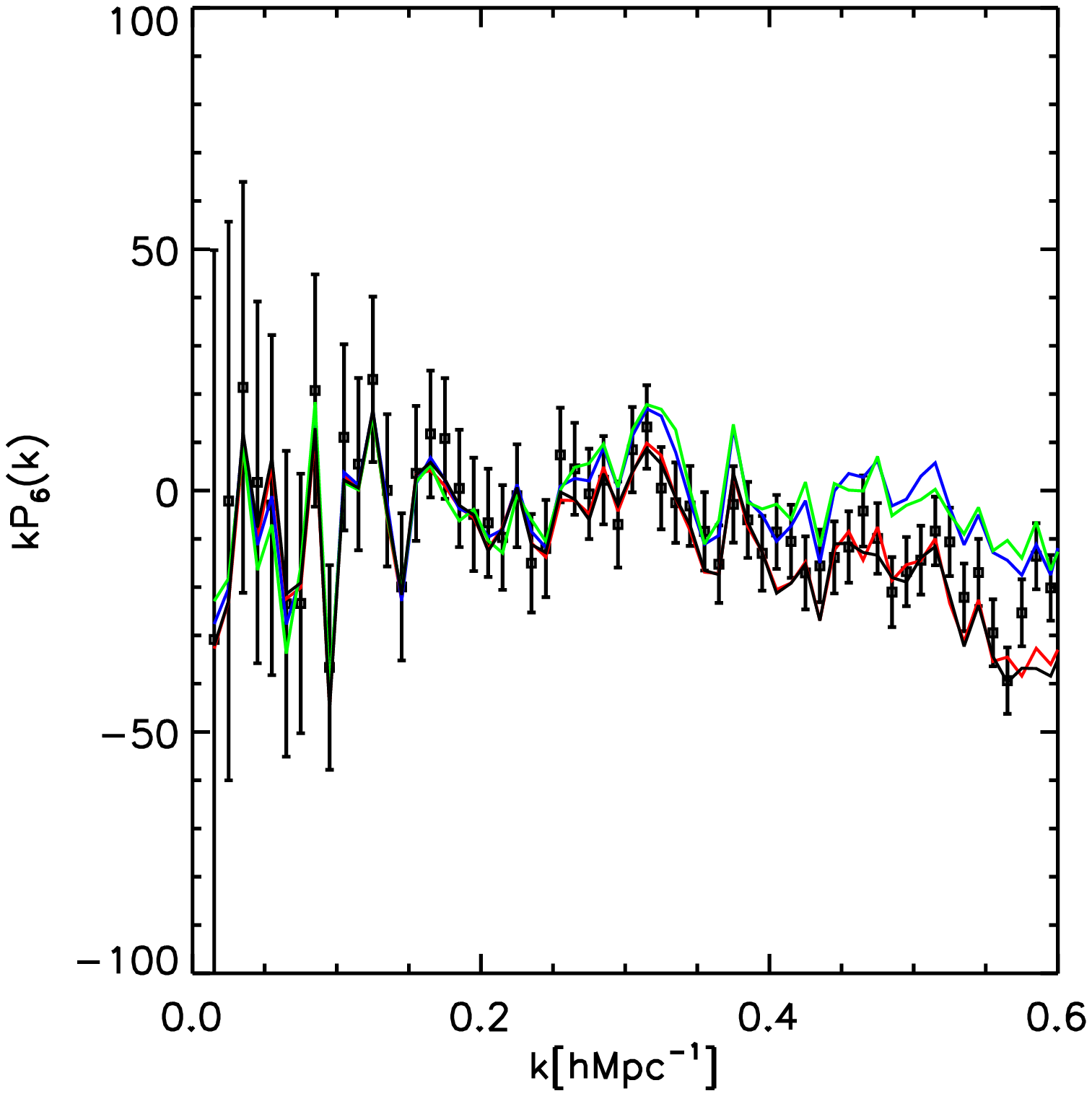}
\caption{Same as Figure~\ref{fig:satellite} but comparing the results
  with the All LRG sample (black curve), the BLRG sample (blue curve),
  and the Single LRG sample (green curve).  The red curve is obtained
  by summing each component of the BLRG sample, the NBLRG sample and
  the cross correlation component. The agreement between the red curve
  and the black curve shows a consistency of the computation. 
\label{fig:contamination}}
\end{center}
\end{figure*}

\subsection{Multipole power spectrum}
\label{sec:satellite}
We adopt the method to measure the multipole power spectrum developed
in \cite{YNKBN}.  For simplicity, we adopt the weight factor $\psi=1$,
and the parameter $\alpha=0.1$ for the random catalog (see
\cite{YNKBN} for details).  The method doesn't take the window effect
of the survey region into account, but it is demonstrated that the
window effect in our method is negligible by comparing with other 
method incorporating it explicitly \cite{sato}.  We perform the
multipole power spectrum analysis for each sample, whose results are
shown in Figure~\ref{fig:satellite} and \ref{fig:contamination}.

Figure~\ref{fig:satellite} compares the multipole power spectrum of
the All LRG sample (black curve) and that of the NBLRGs (red diamond
with large error bars).  The squares with small error bars are the
results in a previous work in \cite{fr}, which are obtained from the
LRG sample with $7150$ square degrees sky coverage with the total
number $100157$ in the range of redshift $0.16\leq z\leq 0.47$.  Thus
the previous sample is almost same as the ``All'' LRG sample in the
present paper.  This figure shows that the amplitude of the
correlation of the NBLRGs is quite large compared with the dominant
component.

Figure~\ref{fig:contamination} shows the effect of the contamination
of the NBLRG on the multipole power spectrum.  The black curve is the
results with the All LRG sample, the green curve is the one with the
Single LRG sample, and the blue curve is the one with the BLRG sample.
The difference between the green and blue is small, which means that
the difference between the Single LRG sample and the BLRG sample is
small.  But the difference between the black curve and the blue curve
is significant, which means that the contribution from the NBLRG
sample is crucial though the fraction of the NBLRGs are small.  This
feature is significant for $P_2(k)$ and $P_4(k)$, especially.  Thus,
the contamination of the satellite galaxy is quite important in these
multipole power spectra.

\begin{figure*}[t]
\includegraphics[scale=.45]{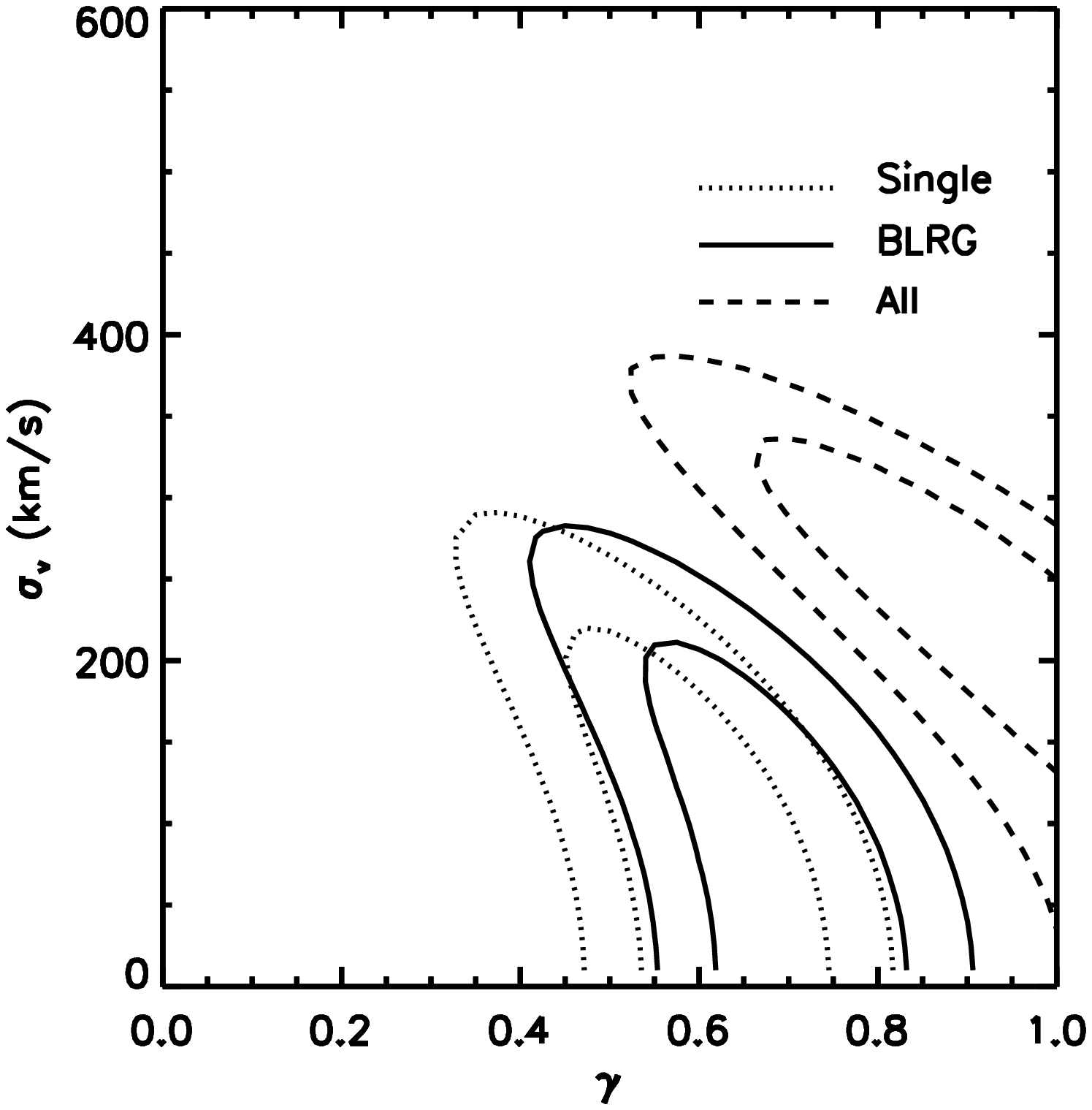}
\includegraphics[scale=.45]{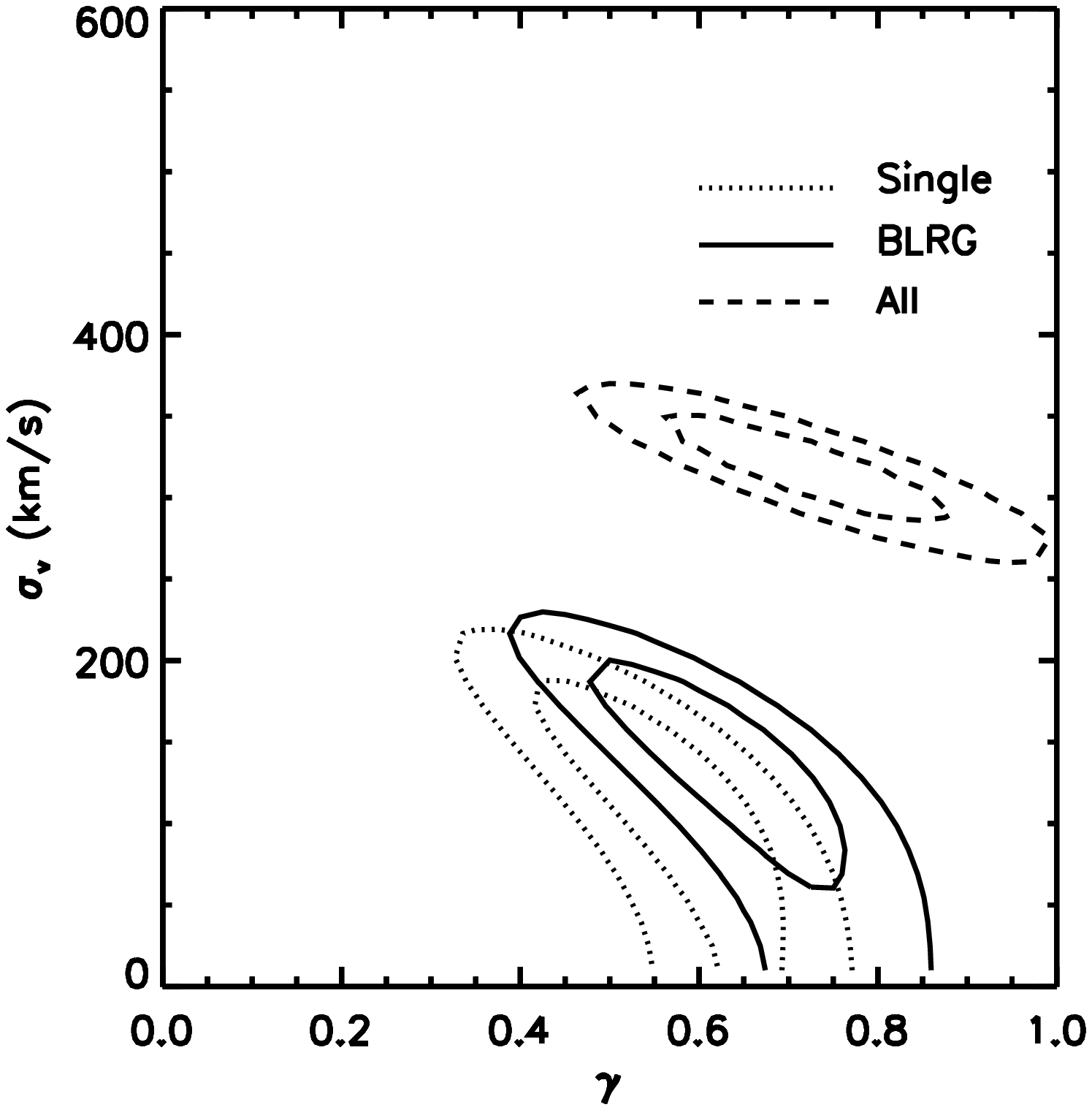}
\caption{Contour of $\Delta \chi^2$ on $\widetilde\sigma_v$ and
  $\gamma$ plane.  The solid (dotted) curves are the 1 sigma and the 2
  sigma contours with the power spectrum with the brightest (single)
  LRG sample, while the dashed curve is the same but with the All LRG
  sample.  The left panel used the data in the range of wavenumbers
  $0.01h{\rm Mpc}^{-1}\leq k\leq 0.2h{\rm Mpc}^{-1}$, but the right
  panel used the data in the range $0.01h{\rm Mpc}^{-1}\leq k\leq
  0.3h{\rm Mpc}^{-1}$.  
\label{fig:impact}}
\end{figure*}

\subsection{Impact on parameter estimation}
\label{sec:parameter}
Here let us demonstrate the impact of the contamination from the
satellite galaxies (NBLRGs) in an estimation of cosmological
parameters.  For simplicity, let us consider the simple model of the
anisotropic power spectrum
\begin{eqnarray}
\label{eq:pkfit}
P(k,\mu)=(b(k)+f\mu^2)^2P_{\rm m}^{\rm NL}(k){\cal D}
\left[k\mu\widetilde\sigma_v/H_0\right],
\end{eqnarray}
where $P_{\rm m}^{\rm NL}(k)$ denotes a nonlinear matter power
spectrum, ${\cal D}[k\mu\widetilde\sigma_v/H_0]$ is 
the damping factor due
to the FoG effect and $\widetilde\sigma_v^2$ is the velocity
dispersion parameter, for which we adopt the function 
\begin{eqnarray}
\label{eq:calD}
{\cal D}[x]={1\over 1+x^2/2}.
\end{eqnarray}
Here we determined the bias $b(k)$ so that the observational and the
theoretical monopole spectra match. Then computed the chi-squared
using the quadrupole spectrum by $\chi^2=\sum_i [P_2^{\rm obs.}(k_i)-
  P_2^{\rm theo.}(k_i)]^2/[\Delta P_2^{\rm obs.}(k_i)]^2$, where
$P_2^{\rm obs.}(k_i)$ and $\Delta P_2^{\rm obs.}(k_i)$, are the
observed values and errors, and $P_2^{\rm theo.}(k_i)$ is the
corresponding theoretical value. See reference~\cite{gr} for details.

Figure~\ref{fig:impact} shows the 1 sigma and 2 sigma contours of
$\Delta\chi^2$ on the parameter plane $\widetilde\sigma_v$ and
$\gamma$, where the growth factor and the growth rate are
parametrized as 
\begin{eqnarray}
&&D_1(a)=a\exp\left[\int_0^a{da'\over a'}(\Omega_m(a')^\gamma-1)\right], \\
&&f(a)={d\log D_1(a)\over d\log a}=\Omega_m(a)^\gamma,
\end{eqnarray}
where $\Omega_m(a)$ is the matter density parameter at the scale
factor $a$.  Here we fixed the other parameters $n_s=0.97$,
$\Omega_m=0.28$, $\Omega_b=0.046$, $\sigma_8=0.8$ and assumed the cold
dark matter model with a cosmological constant ($\Lambda$CDM model) as the background
universe model.  In each panel, the dotted curve, solid curve, and the
dashed curve are the Single, Brightest, and All LRG sample,
respectively. The left (right) panel used the data with $k\le 0.2h{\rm
  Mpc}^{-1}$ ($k\le 0.3h{\rm Mpc}^{-1}$).  The value $\gamma=0.55$ is
the prediction of the model on the basis of the general relativity
\cite{Linder}.  Though our theoretical model is very simple, the
results clearly show that the contamination of the satellite galaxies
(NBLRGs) significantly biases the parameter estimation.  This figure
also indicates that the results are influenced by including the
brightest LRGs consisting of the multiple systems.

\section{Halo model description of satellite Finger-of-God}
\label{sec:an}
In this section, we consider the FoG effect of satellite galaxies
based on the halo model picture
\cite{Seljak2001,White2001,CooraySheth2002}.  In the halo model, the
power spectrum of LRGs are decomposed into 1-halo and 2-halo
terms. Then we write the anisotropic power spectrum in the
redshift-space consisting of the 1-halo and 2-halo terms,
\begin{equation}
P_{\rm LRG}(k,\mu)=P^{\rm 1h}(k,\mu)+P^{\rm 2h}(k,\mu).
\end{equation}
We here consider the sample which consists of the central galaxies and
the satellite galaxies, and adopt the following expressions
(\ref{eq:pk_1h}) and (\ref{eq:pk_2h}) for $P^{\rm 1h}(k,\mu)$ and
$P^{\rm 2h}(k,\mu)$, respectively.  A brief summary of the derivation
for a general case is described in the appendix (See also below for
details).

\begin{figure}[t]
\begin{center}
\includegraphics[width=8cm]{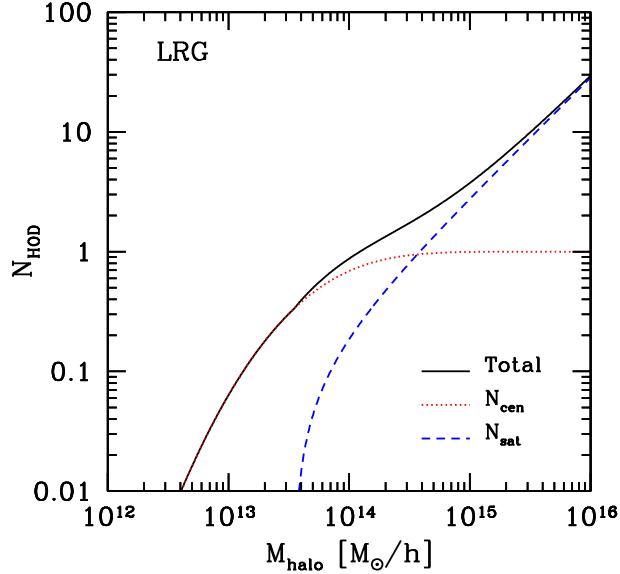} 
\caption{HOD for LRGs based on \cite{Reid2009a}.
\label{fig:hod_lrg}}
\end{center}
\end{figure}

One-halo term is given by
\begin{eqnarray}
P^{\rm 1h}(k,\mu)=\frac{\displaystyle 1}{\displaystyle \bar{n}^2}\int\!dM~
\frac{\displaystyle dn}{\displaystyle dM}
\Bigl[2\langle N_{\rm cen}\rangle \langle N_{\rm sat}\rangle
\tilde{p}_{\rm cs}(k,\mu; M)
+\langle N_{\rm sat}(N_{\rm sat}-1)\rangle
\tilde{p}_{\rm ss}(k,\mu; M)\Bigr],
\label{eq:pk_1h}
\end{eqnarray}
where we adopt the halo mass function $dn/dM$ given by
\cite{ShethTormen1999} and $\bar{n}$ is the mean number density of
LRGs given by $\bar{n}=\int dM (dn/dM) N_{\rm HOD}(M)$ and $N_{\rm
  HOD}(M)$ is the halo occupation distribution (i.e., the average
number of galaxies inside the halo with mass $M$). We use the
following form of the HOD of central LRGs and satellite LRGs
\cite{Zheng2005}
\begin{eqnarray}
&&N_{\rm HOD}(M)=\langle N_{\rm cen}\rangle(1+\langle N_{\rm sat}\rangle), \\
&&\langle N_{\rm cen}\rangle =\frac{1}{2}\left[1+{\rm erf}\left(\frac{\log_{10}(M)-\log_{10}
(M_{\rm min})}{\sigma_{\log M}}\right)\right], \\
&&\langle N_{\rm sat}\rangle =f_{\rm col}(M)
\left(\frac{M-M_{\rm cut}}{M_1}\right)^{\alpha},
\label{eq:HOD}
\end{eqnarray}
where ${\rm erf}(x)$ is the error function. We adopt $M_{\rm
  min}=5.7\times 10^{13}M_\odot/h$, $\sigma_{\log M}=0.7$, $M_{\rm
  cut}=3.5\times 10^{13}M_\odot/h$, $M_1=3.5\times 10^{14}M_\odot/h$,
and $\alpha=1$ to match the HOD of SDSS DR7 LRG catalog
\cite{Reid2009a} as shown in Figure~\ref{fig:hod_lrg}.  
Assuming the number of groups with $N_{\rm sat}$ satellites is Poisson
distributed \cite{Kravtsov2004}, the averaged satellite-satellite pair
number $\langle N_{\rm sat}(N_{\rm sat}-1)\rangle$ per halo goes to
$\langle N_{\rm cen}\rangle\langle N_{\rm sat}\rangle^2$.  
We also take into account the missing galaxies due to the fiber collision by
multiplying the satellite HOD with a following mass-dependent factor
\begin{equation}
f_{\rm col}(M)=A_{\rm col}+B_{\rm col}\left(\frac{M-M_{\rm cut}}{M_1}\right),
\end{equation}
where $1-f_{\rm col}(M)$ represent the fraction of missing satellite LRGs due to
the fiber collision effect for the host halo mass of $M$.  The
factor $A_{\rm col}$ and $A_{\rm col}+B_{\rm col}$ corresponds to
$f_{\rm col}(M)$ for $M=M_{\rm cut}$ and $M=M_1$ where the averaged
number of satellites is 0 and 1 respectively.  Here we set $A_{\rm
  col}=0.7$ and $B_{\rm col}=-0.05$ to match the number fraction of
NBLRGs and the number of NBLRG pairs in groups.  We do not consider
the fiber collision effect on central HOD, for simplicity.

Central LRGs locate near the halo center and thus their velocity
difference relative to the host halo should be small.  Note that it is
difficult to verify that each central LRG is located at the center of
each halo in observational data.  However, 20-40\% of brightest LRGs
are found to be off-centered (satellite) galaxies using lensing and
cross-correlation analysis \cite{Hikage2012}. Therefore, large part of
the NBLRGs are off-centered and their velocity should be the main
source of the FoG effect.  The functions $\tilde p_{\rm cs}(k,\mu;M)$
and $\tilde p_{\rm ss}(k,\mu;M)$ are the Fourier transform of
central-satellite and satellite-satellite distribution inside the halo
with the mass of $M$, and the internal motion of satellite LRGs
elongate the distributions in the line-of-sight direction. We assume
that the internal velocity of the satellite LRGs has a Gaussian
distribution determined by virial velocity as $\sigma_{v,{\rm
    off}}(M)=(GM/2R_{\rm vir})^{1/2}$, in which the virial radius of
the halo with mass of $M$ is $R_{\rm vir}=(3M/4\pi\bar{\rho}_{\rm
  m}(z)\Delta_{\rm vir}(z))^{1/3}$ with $\Delta_{\rm vir}=265$ at
$z=0.32$.  When the satellite motion is uncorrelated with each other,
$\tilde p_{\rm cs}(k,\mu,M)$ and $\tilde p_{\rm ss}(k,\mu,M)$ are
given by
\begin{eqnarray}
\label{eq:psoff}
\tilde{p}_{\rm cs}(k,\mu,M)&=&\tilde{u}_{\rm NFW}(k;M)\exp\left[-\frac{\sigma_{v,{\rm off}}^2(M)k^2\mu^2}{2a^2H^2(z)}\right], \\
\tilde{p}_{\rm ss}(k,\mu,M)&=&\tilde{p}_{\rm cs}^2(k,\mu,M).
\label{eq:psoff2}
\end{eqnarray}
We assume that the distribution of the satellite galaxies follows 
the NFW profile \cite{NFW1996} and $\tilde{u}_{\rm NFW}(k)$ denotes 
the Fourier transform of truncated NFW profile, equation~(\ref{tuNFW}),  
(see also \cite{Scoccimarro2001}).
\begin{figure*}
\begin{center}
\includegraphics[width=16cm]{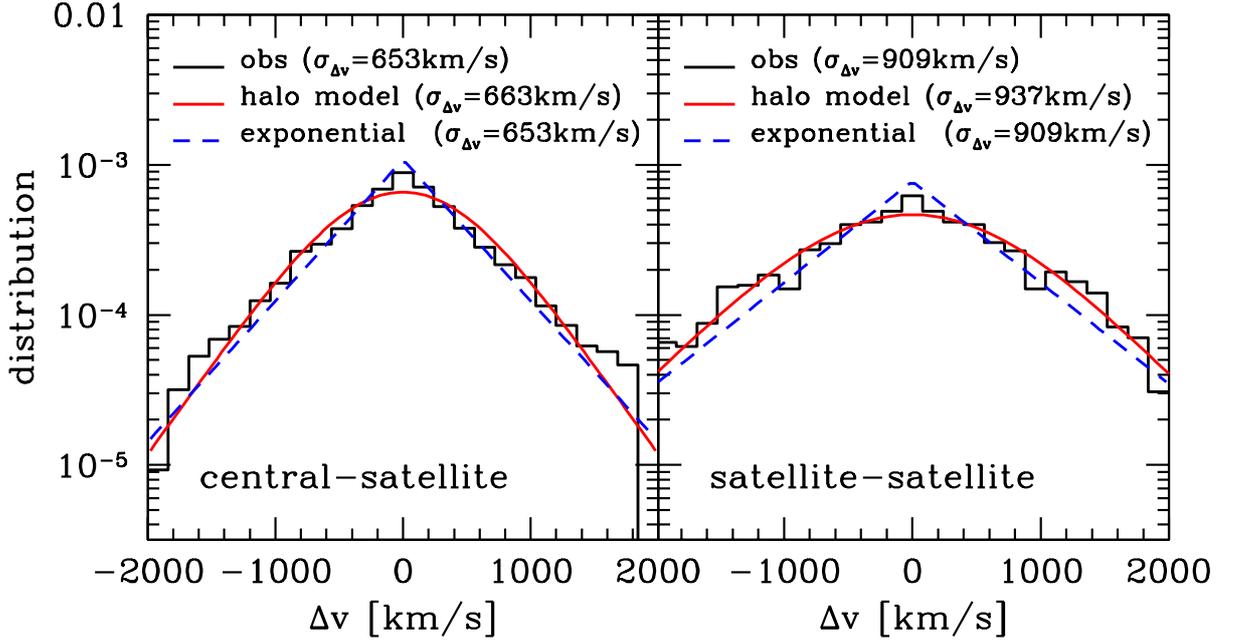}
\caption{Pairwise velocity distribution between
    central-satellite (left) and satellite-satellite
    (right). Histogram indicates the observed pairwise velocity
    distribution obtained from the redshift differences between BLRGs
    and NBLRGs (left) and among NBLRGs (right) in the same groups. The
    value of $\sigma_{\Delta v}$ represent the r.m.s of the averaged
    pairwise velocity dispersion.  For comparison, we plot the
    theoretical predictions based on the halo model (solid red
    curves), equation~(\ref{eq:pairvel_cs}) in the left panel and
    equation~(\ref{eq:pairvel_ss}) in the right panel, and an
    exponential profile with the dispersion of observed value of
    $\sigma_{\Delta v}$ (blue dashed curves).
\label{fig:veldif}}
\end{center}
\end{figure*}
In order to test the validity of Gaussian assumption of satellite
velocity distribution, equations~(\ref{eq:psoff}) and (\ref{eq:psoff2}), we compare
the distribution functions of pairwise velocity for central-satellite pairs and
satellite-satellite pairs based on the halo model, as shown in
Figure \ref{fig:veldif}. We compute the pairwise velocity between
NBLRGs and BLRGs within the same group from their redshift difference as
$\Delta v=c\Delta z/(1+z)$. We find that the distributions are well
explained by the mass integral of the Gaussian velocity distribution
with the Virial velocity dispersion of each mass $\sigma_{v,{\rm
    off}}(M)=(GM/2R_{\rm vir})^{1/2}$,
\begin{eqnarray}
P(\Delta v)^{\rm cen-sat}&\propto &\int dM\frac{dn}{dM}\langle N_{\rm cen}\rangle
\langle N_{\rm sat}\rangle \exp\left(-\frac{\Delta v^2}{2\sigma_{v,
    {\rm off}}^2(M)}\right), 
\label{eq:pairvel_cs}
\\
P(\Delta v)^{\rm sat-sat}&\propto &\int dM\frac{dn}{dM}\langle N_{\rm sat}(N_{\rm sat}-1)\rangle
\exp\left(-\frac{\Delta v^2}{4\sigma_{v,{\rm off}}^2(M)}\right).
\label{eq:pairvel_ss}
\end{eqnarray}
where the normalization of $P(\Delta v)$ is determined so that the
integral over $\Delta v$ is unity.  With the velocity probability
distribution functions, we compute $\sigma_{\Delta v}^2=\int\Delta v^2
P(\Delta v) d\Delta v$ for the theoretical value of pairwise velocity
dispersions. The model predictions become 663km/s for central-satellite
pairs and 937km/s for satellite-satellite pairs. These values well
agree with the observed pairwise velocity dispersions: 653km/s for
BLRG-NBLRG pairs and 909km/s for NBLRG-NBLRG pairs. The good agreement
validates our models of central-satellite and satellite-satellite
distributions in redshift space (equations~(\ref{eq:psoff}) and
(\ref{eq:psoff2})).  For comparison, we also plot the exponential
profile, which also well describes the behavior of the observed
pairwise velocity distribution.

The 2-halo term is given by
\begin{eqnarray}
P^{\rm 2h}(k,\mu)=\biggl[\frac{\displaystyle 1}{\displaystyle \bar{n}}
\int\!dM~\frac{\displaystyle dn}{\displaystyle dM}(b(M)+f\mu^2)
\langle N_{\rm cen}\rangle~~~~~~~~~~~~~~~~ \nonumber \\
\times
(1+\langle N_{\rm sat}\rangle\tilde{p}_{\rm cs}(k,\mu; M))\tilde{u}_{\rm vol}(k;M)
\biggr]^2P^{\rm NL}_{\rm m}(k), 
\label{eq:pk_2h}
\end{eqnarray}
where $P^{\rm NL}_{\rm m}(k)$ is the real-space non-linear matter
power spectrum. Here we use the non-linear matter power spectrum
  to describe the non-linear power spectrum of velocity divergence for
  simplicity, while the matter and velocity power spectra are actually
  different (c.f.,\cite{Scoccimarro2004,Matsubara2008}).  Here we add the volume
exclusion effect of halos $\tilde{u}_{\rm vol}(k;M)$ in addition to
the satellite distribution in order to include that two different
halos cannot approach each other closer than a halo size.  We use a
Gaussian form $\tilde{u}_{\rm vol}(k;M)=\exp(-(akR_{\rm vir}(M))^2/2)$
and we choose the width parameter $a=2$ to fit the observed power
spectrum.

Again we consider only the velocity distribution of satellite LRGs. We
simply use the linear Kaiser formula \cite{Kaiser1987} given by the
term of $(b(M)+f\mu^2)$ with the growth rate $f\equiv d\ln D/d\ln a$
and the linear halo bias $b(M)$~\cite{MoJingWhite1997,Scoccimarro2001,CooraySheth2002}.
Without the FoG effect, that is $\sigma_{v,{\rm off}}=0$, we have
$P^{\rm 2h}\simeq (b_{\rm eff}+f\mu^2)^2P_{\rm NL}$ and $P^{\rm
  1h}\simeq N_{\rm 1h}$, where $b_{\rm eff}$ is the effective bias of
LRGs given by $b_{\rm eff}=\int dM (dn/dM) b(M) N_{\rm
  HOD}(M)/\bar{n}$ and $N_{\rm 1h}$ is defined by $N_{\rm
  1h}=\int\!dM(dn/dM) \bigl[2\langle N_{\rm cen}\rangle \langle N_{\rm
    sat}\rangle +\langle N_{\rm sat}(N_{\rm sat}-1)\rangle
  \bigr]/{\bar n}^2$. In this case, $N_{\rm 1h}$ is a constant and we
have $P^{1h}_\ell(k)=0$ for $\ell\geq2$.

\begin{figure*}
\includegraphics[scale=.4]{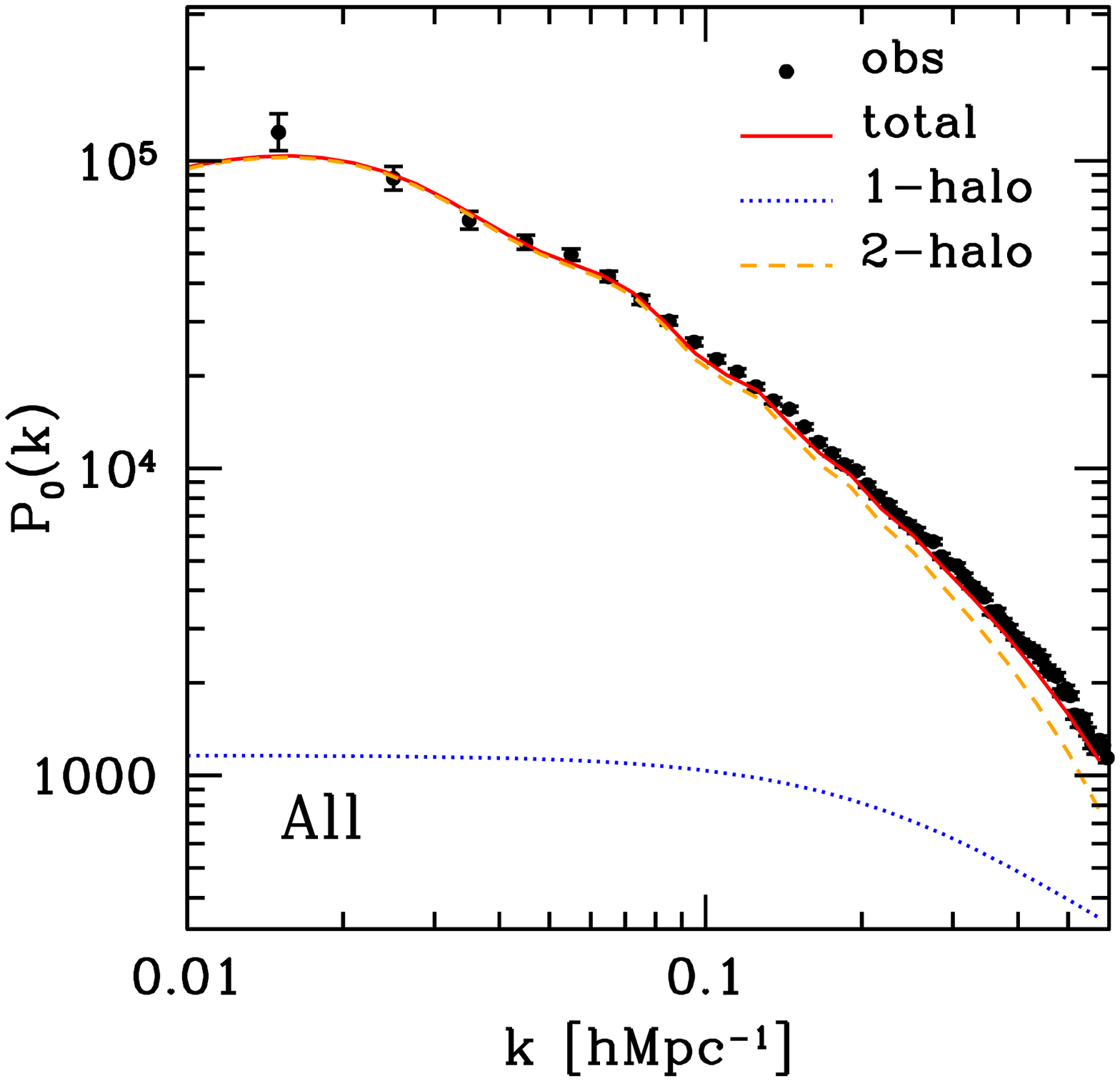}
\includegraphics[scale=.4]{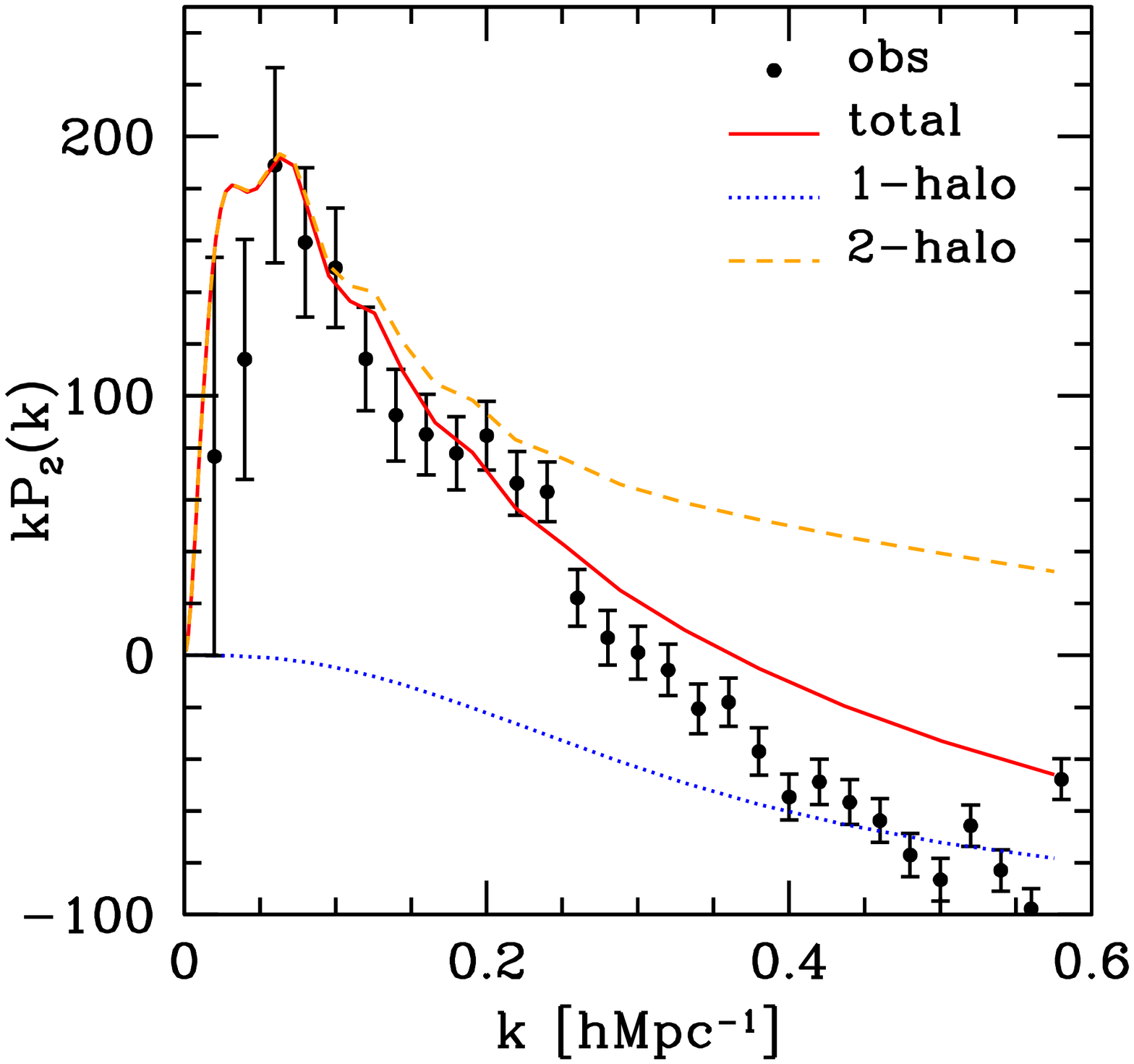}
\includegraphics[scale=.4]{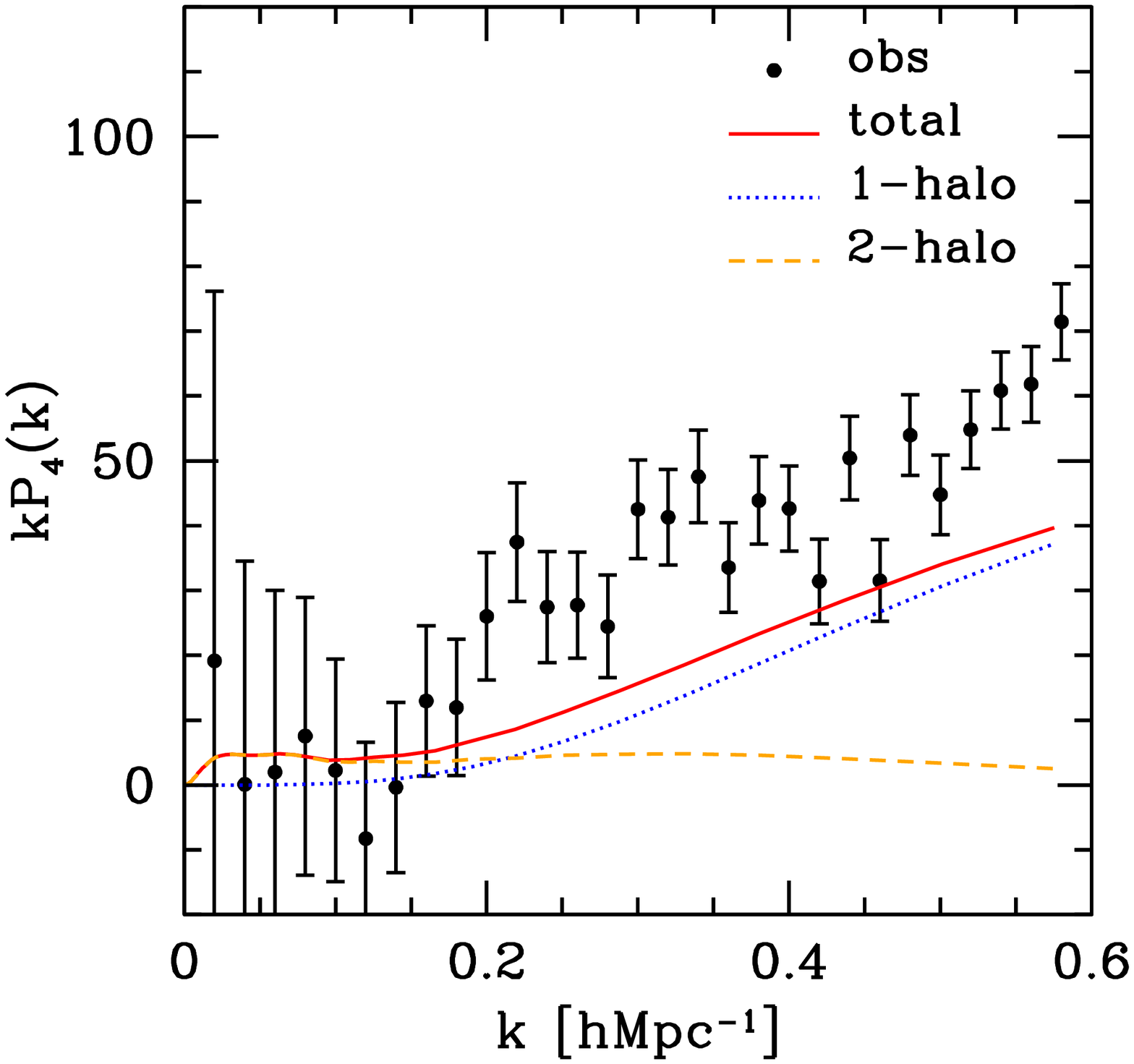}
\includegraphics[scale=.4]{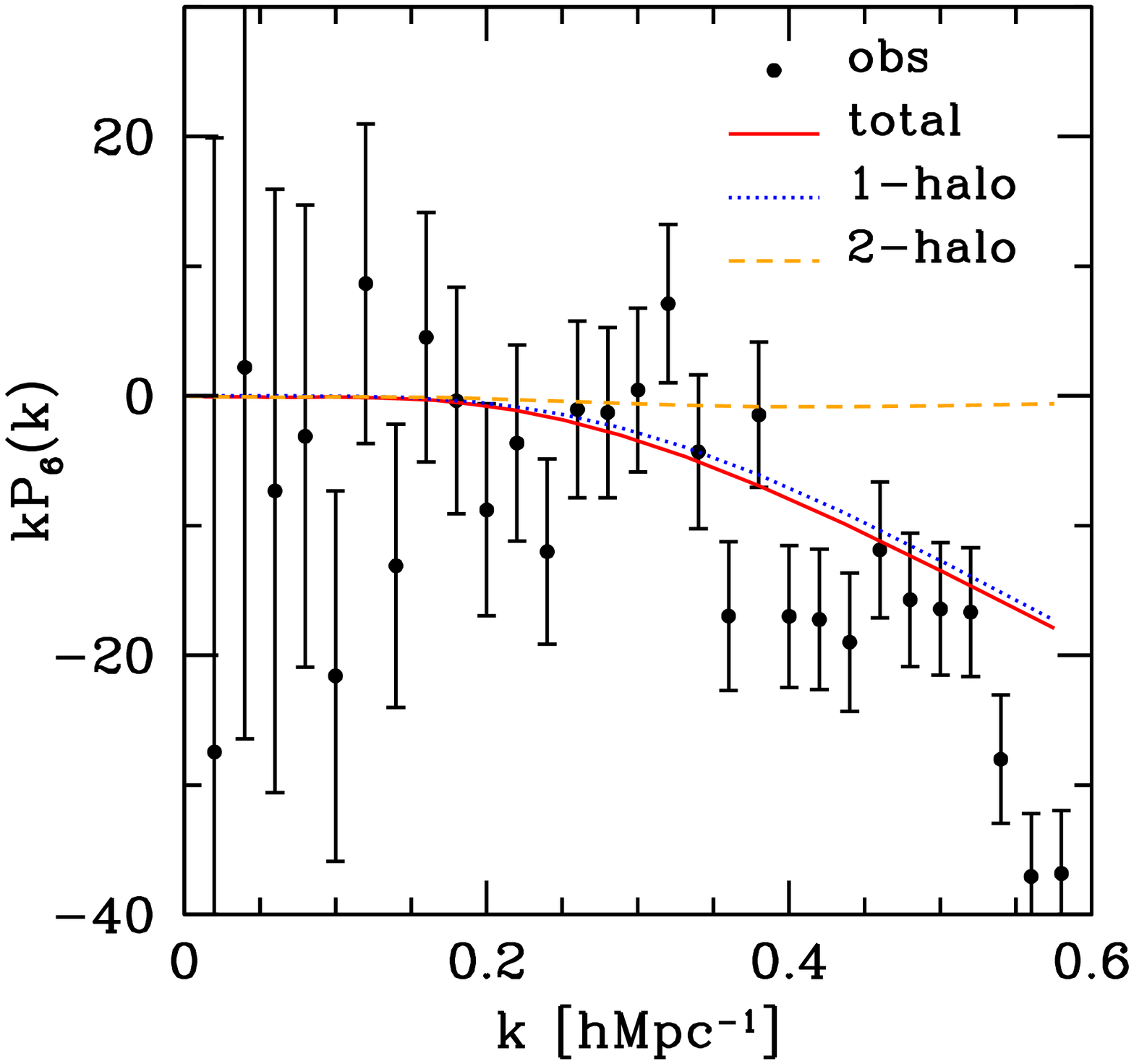}
\caption{Halo model prediction for the multipole power spectra
  $P_0(k)$, $P_2(k)$, $P_4(k)$, and $P_6(k)$ for the All LRG sample.
  In each panel, the dotted curve and the dashed curve are the 1-halo
  term and the 2-halo term, respectively, and the solid curve is their
  combination.  The black circles are the observational data of the
  All LRG sample in Figure~\ref{fig:contamination}. 
\label{fig:model_all} }
\end{figure*}

\begin{figure*}
\includegraphics[scale=.4]{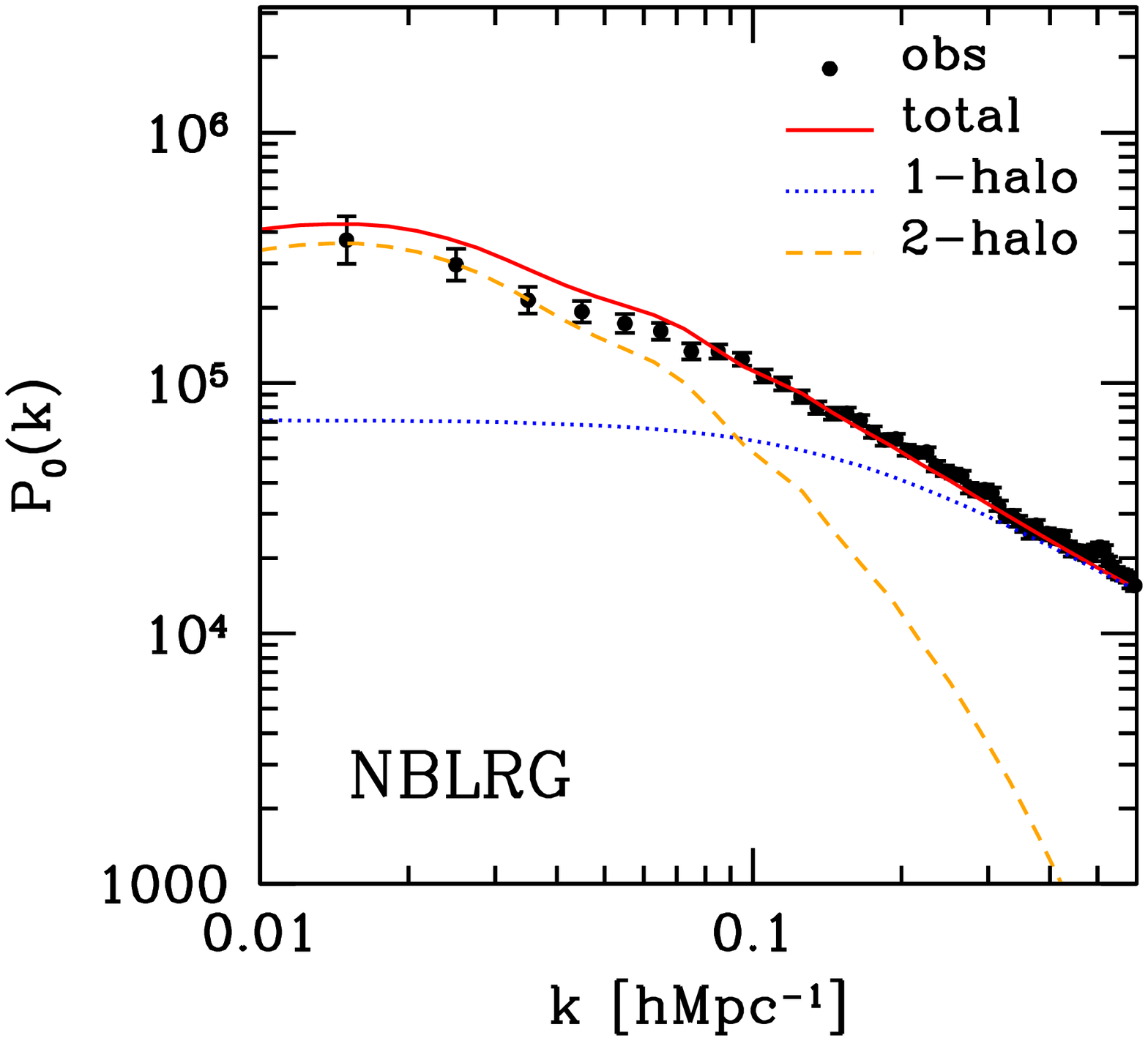}
\includegraphics[scale=.4]{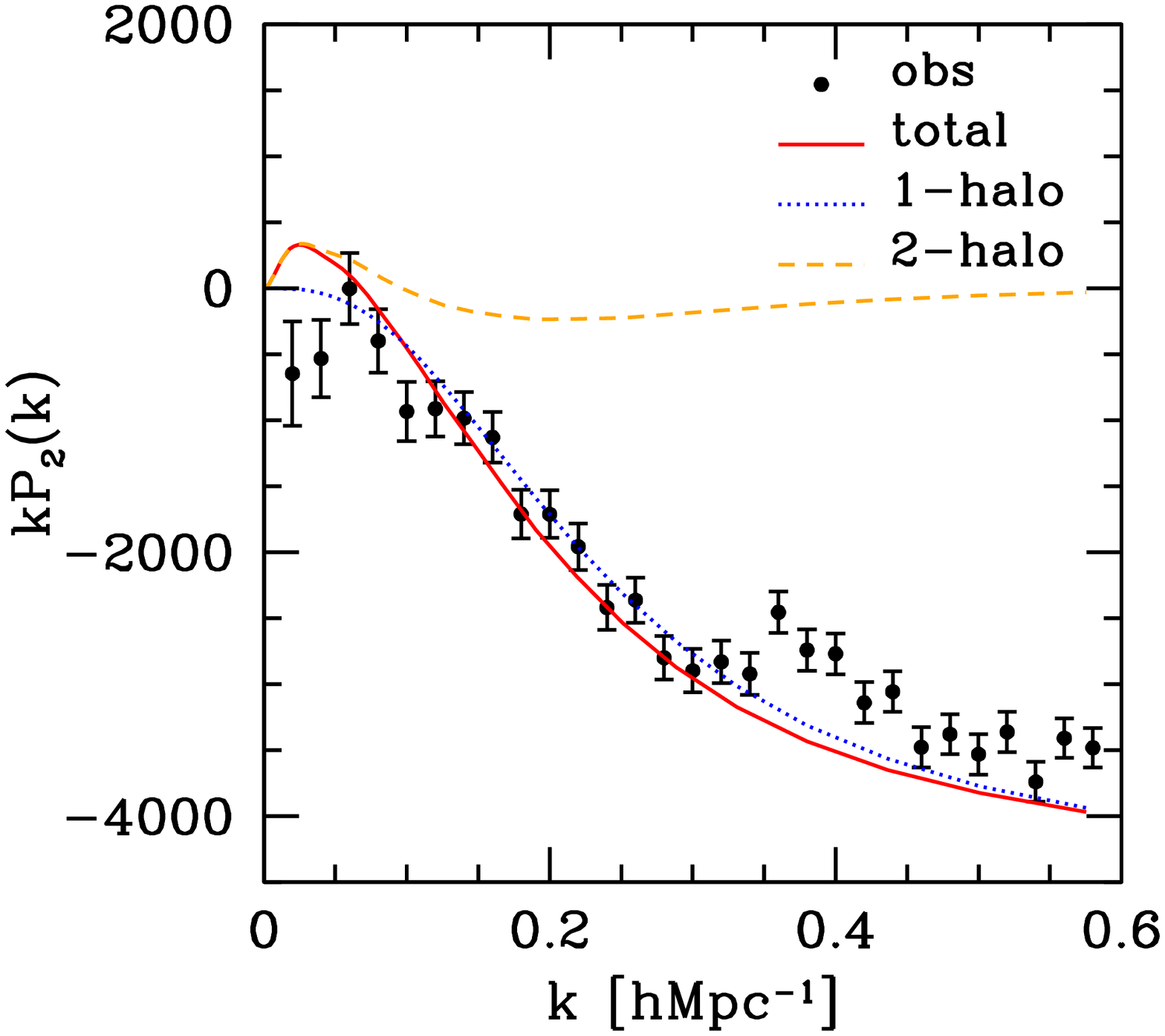}
\includegraphics[scale=.4]{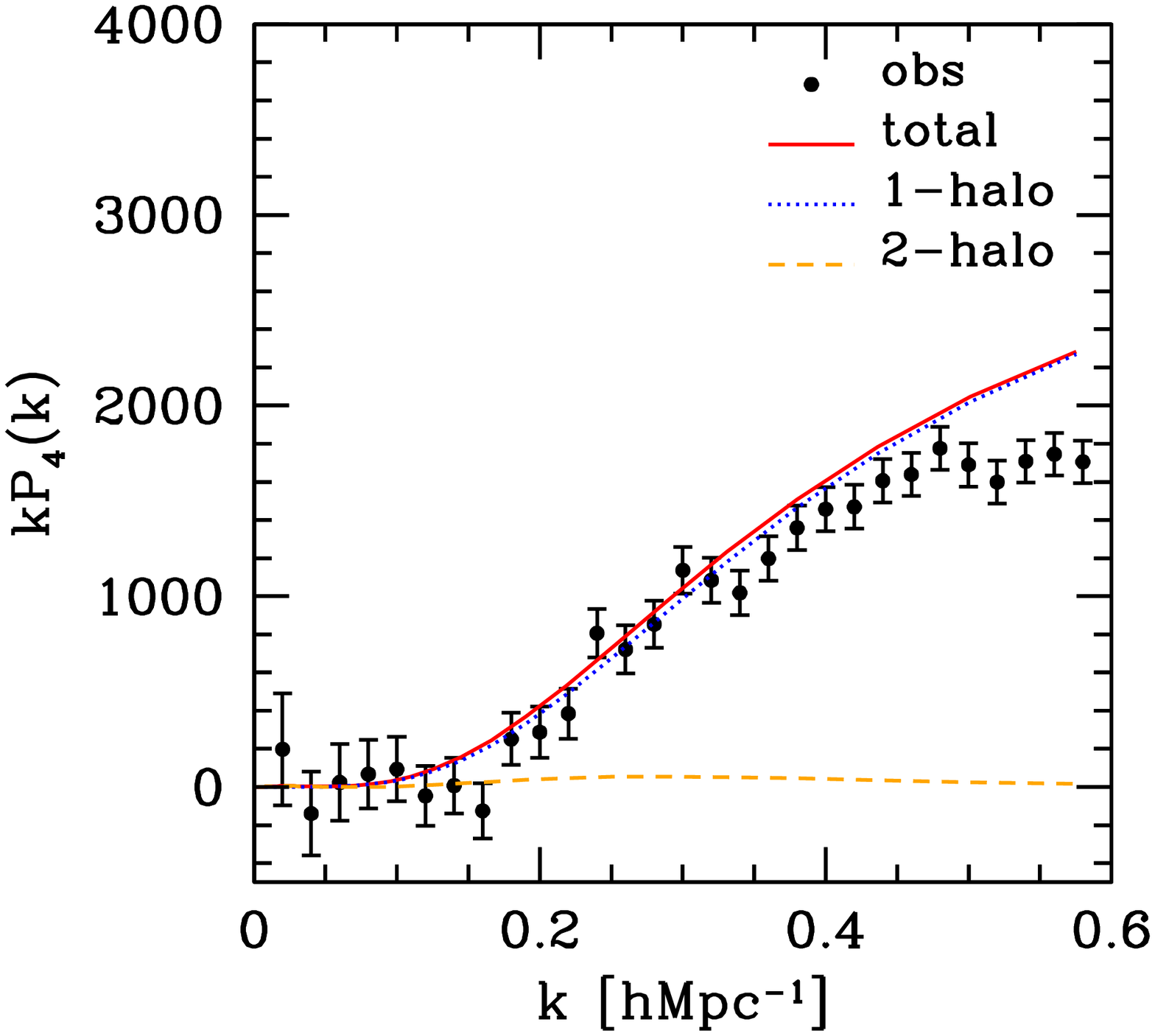}
\includegraphics[scale=.4]{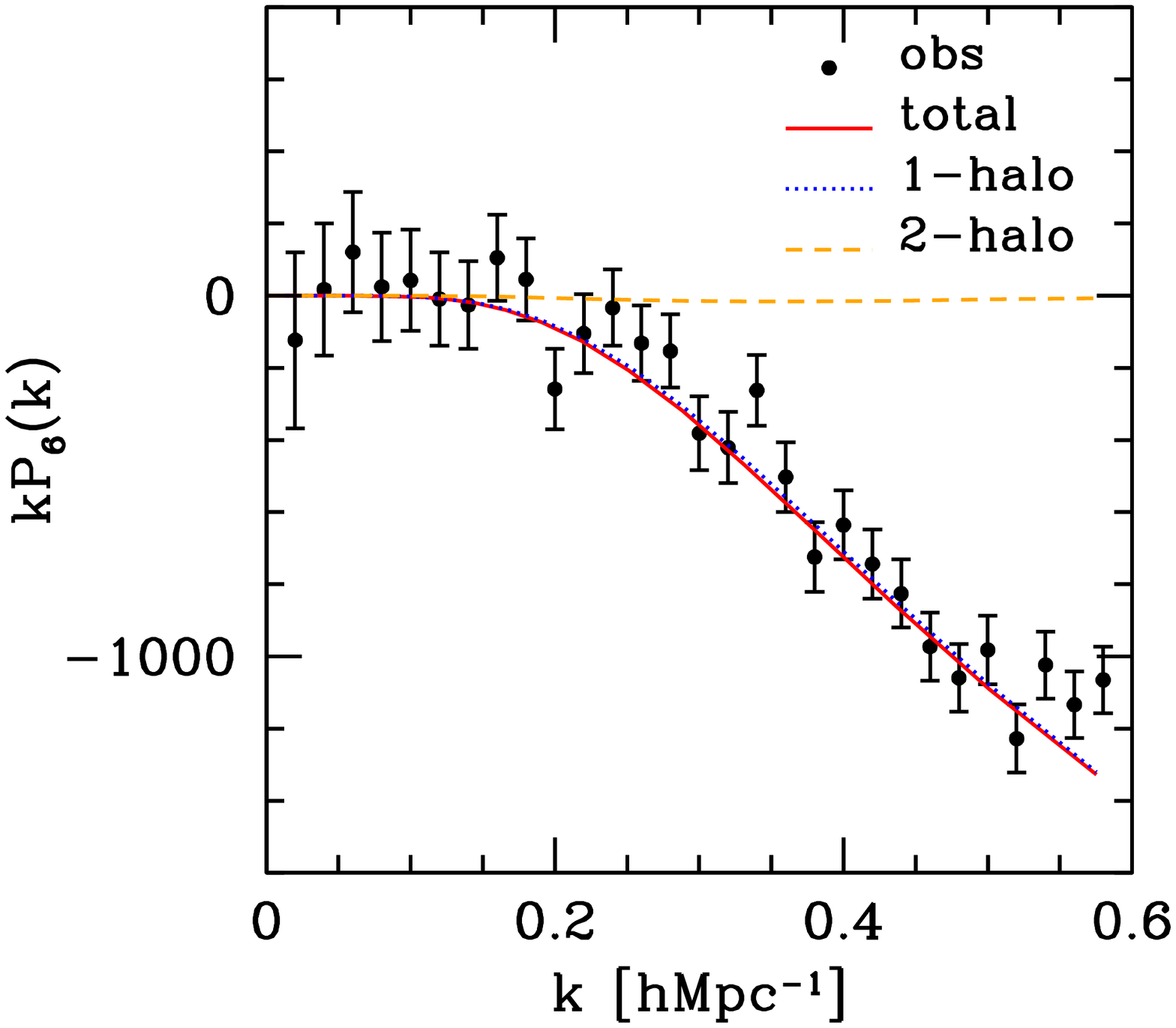}
\caption{Halo model prediction for the multipole power spectra
  $P_0(k)$, $P_2(k)$, $P_4(k)$, and $P_6(k)$ for the NBLRG sample.  In
  each panel, the dotted curve and the dashed curve are the 1-halo
  term and the 2-halo term, respectively, and the solid curve is their
  combination.  Here the BLRG sample is assumed to be consisting of
  the central galaxies (35\%) and the satellite galaxies (65\%).  The
  black circles are the observational data of the NBLRG sample in
  Figure~\ref{fig:satellite}. 
\label{fig:model_sat} }
\end{figure*}

\begin{figure*}[h]
\includegraphics[scale=.4]{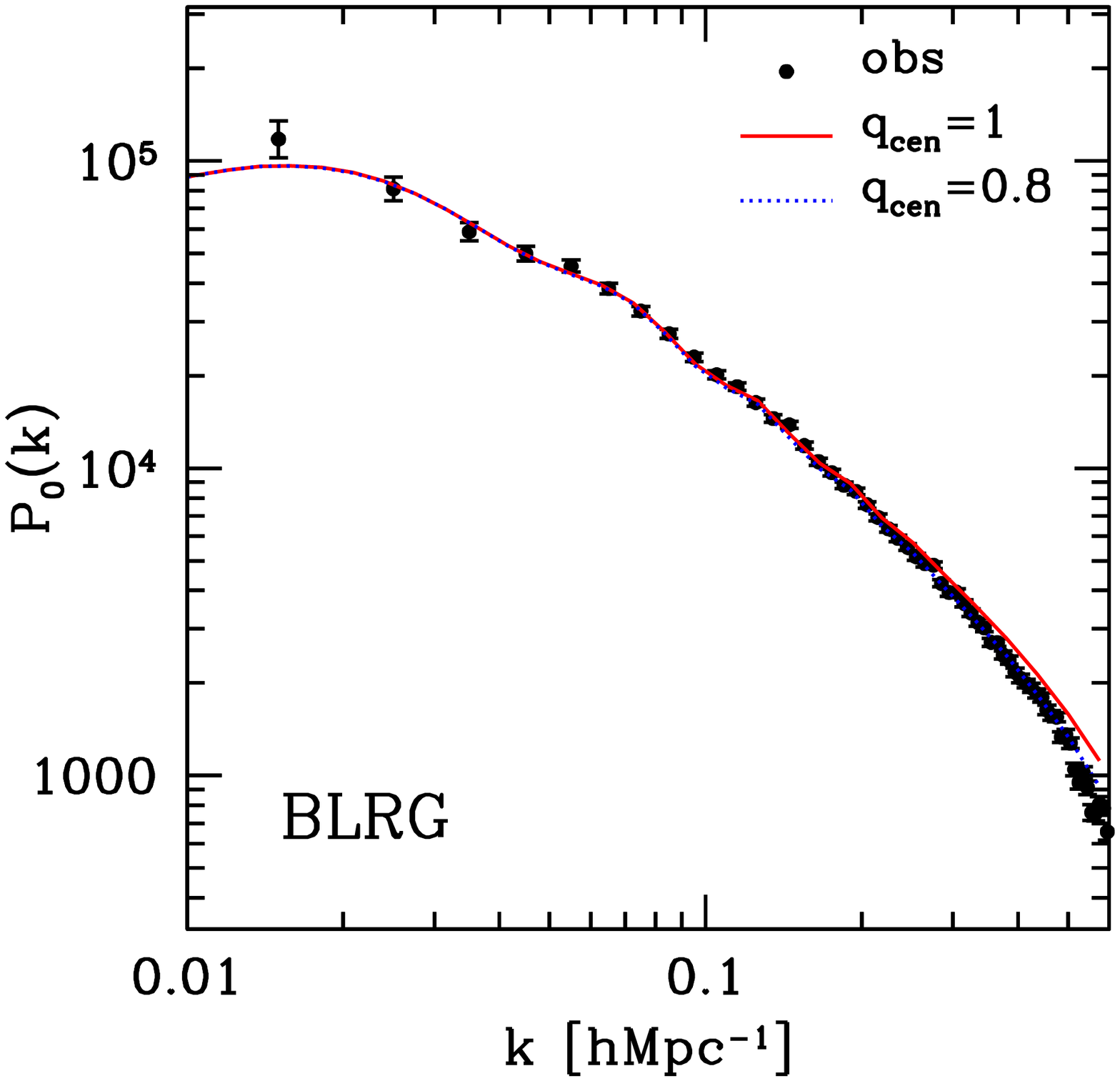}
\includegraphics[scale=.4]{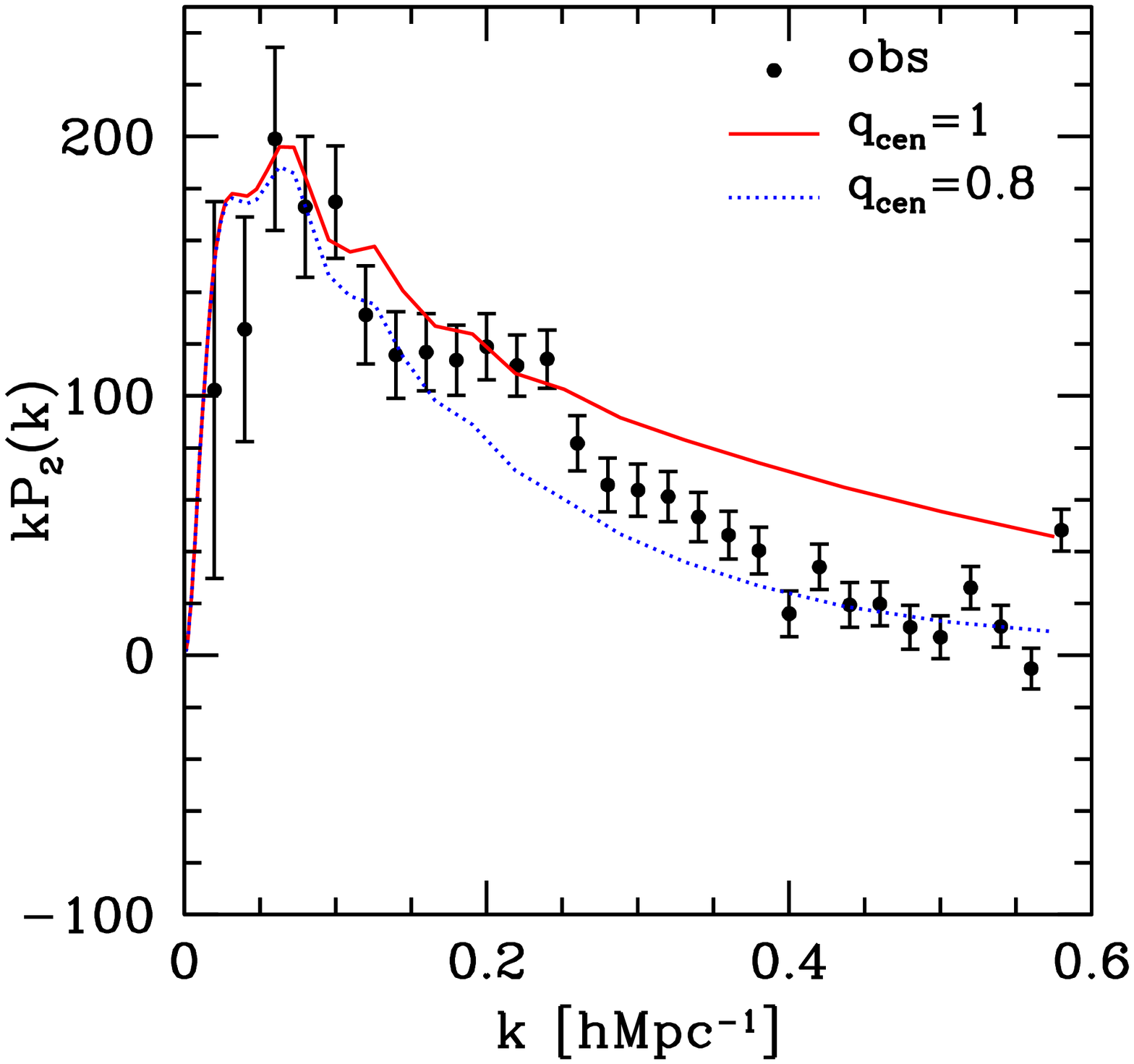}
\includegraphics[scale=.4]{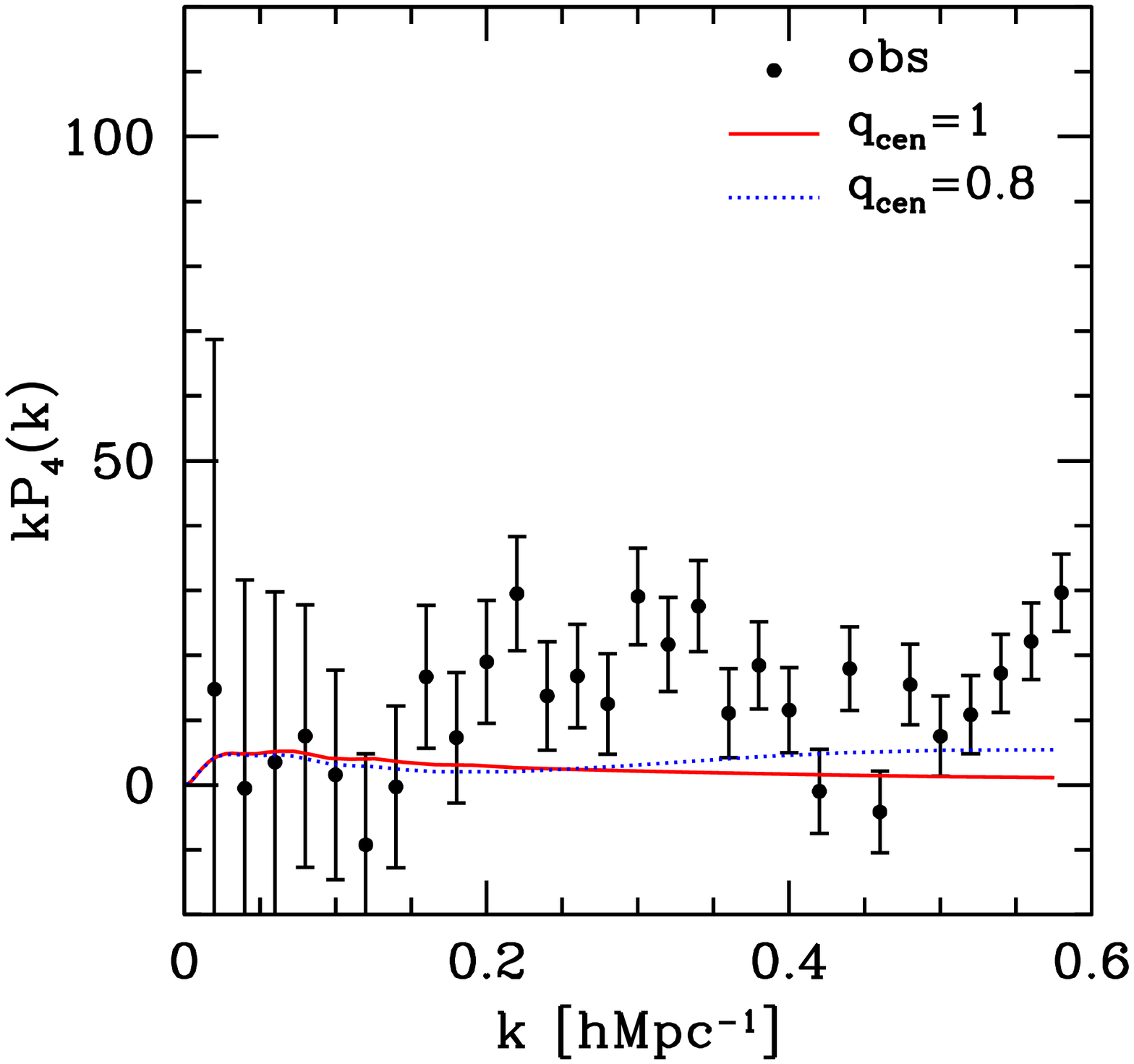}
\includegraphics[scale=.4]{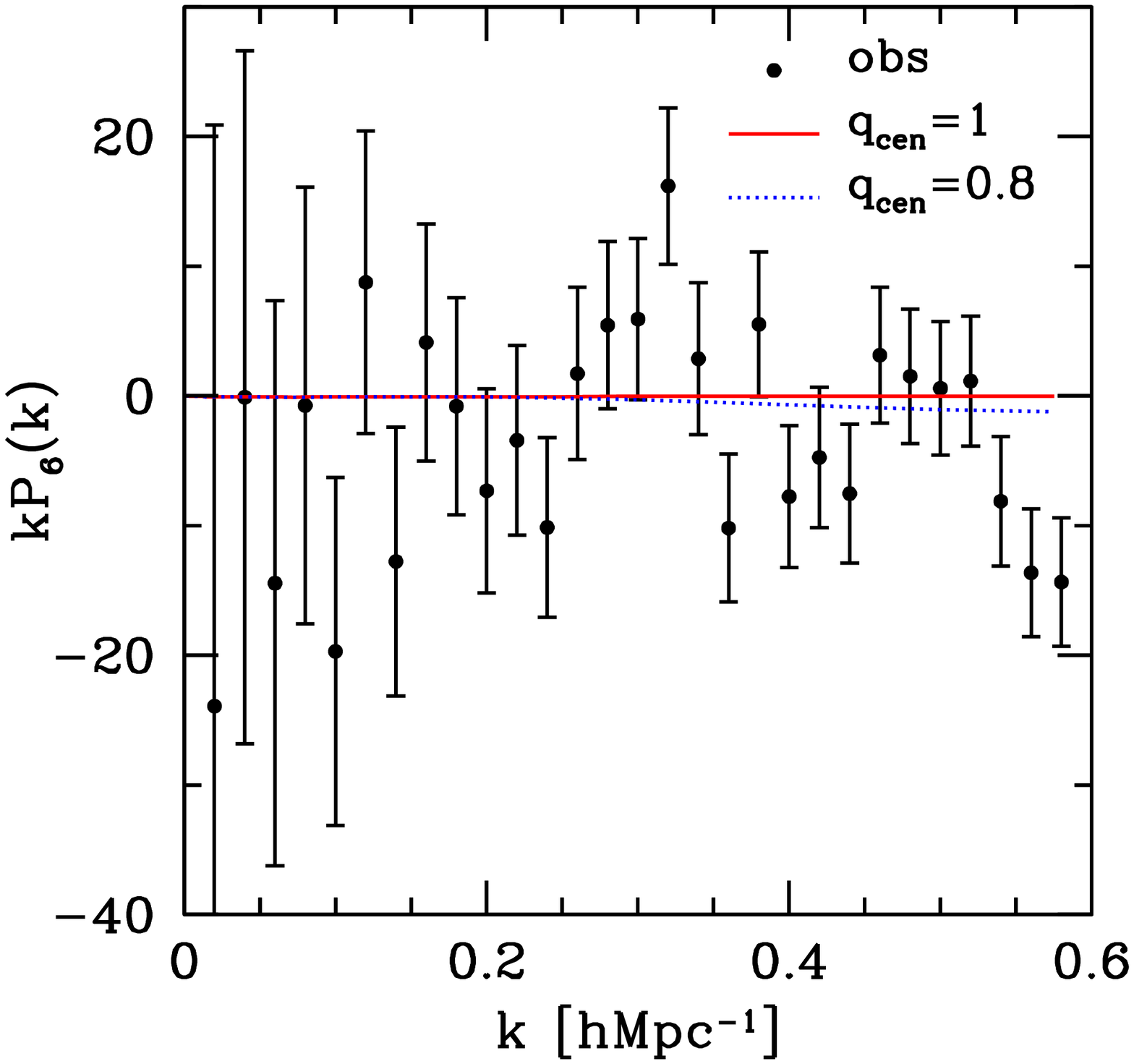}
\caption{ Halo model prediction for the multipole power spectra
  $P_0(k)$, $P_2(k)$, $P_4(k)$, and $P_6(k)$ for the BLRG sample,
  which is written with just 2-halo term.  Here the solid curve
  adopted the fraction of the central galaxies $q_{\rm cen}^{\rm
    (BLRG)}=1$, while the dotted curve did $q_{\rm cen}^{\rm
    (BLRG)}=0.8$.
\label{fig:model_blrg}}
\end{figure*}

Figure~\ref{fig:model_all} compares the halo model predictions of
multipole power spectra for All LRG sample with the observations. Our
model qualitatively well explain the observations although we simply
adopted the linear Kaiser redshift distortion and the linear halo
bias.  The halo model well explains the differences between ALL and
BLRG samples as shown in the below of this section.  The satellite FoG
effect in 1-halo term becomes significantly important at larger $k$
and dominant in the multipole spectra for $\ell\ge 4$ even though the
satellite fraction is only $5$\%. The 1-halo term contribution causes
a systematic bias in the measurement of the growth rate as shown in
Figure~\ref{fig:impact} because the FoG effects from the 1 and 2-halo
terms have different feature and the simple form of equation
(\ref{eq:pkfit}) is not enough to describe both of the FoG effects very well.
In the following section, we show how the constraints on the growth rate
changes by taking into account the 1-halo term.

The behavior of higher-order multipole spectrum is sensitive to the
satellite FoG effect in one-halo term. In other words, the higher
multipole spectra can be a good probe of the satellite fraction and
the satellite velocity distribution.
\def\Erf{{\rm erf}}
The one halo term making contribution to the multipole power spectrum 
of (\ref{eq:pk_1h}) can be written as follows:
\begin{eqnarray}
P^{1h}_\ell(k)={1\over {\bar n}^2}\int dM {dn\over dM}\left[
2\langle N_{\rm cen}\rangle \langle N_{\rm sat}\rangle Q_\ell(q)
+\langle N_{\rm sat}(N_{\rm sat}-1)\rangle Q_\ell(\sqrt2 q)
\right],
\label{P1hlexpress}
\end{eqnarray}
where we defined 
\begin{eqnarray}
 Q_\ell(q)={1\over 2}\int_{-1}^1 d\mu e^{-q^2\mu^2}{\cal L}_\ell(\mu)
\end{eqnarray}
and $q={\sigma_{v,{\rm off}}(M)k/\sqrt{2}aH(z)}$. 
Specifically, we have
\begin{eqnarray}
Q_0(q)&=&\frac{\displaystyle\sqrt{\pi}}{\displaystyle 2q}\Erf(q), 
\label{Q0express}
\\
Q_2(q)&=&-\frac{\displaystyle 3}{\displaystyle 4q^2}e^{-q^2}
+\frac{\displaystyle\sqrt{\pi}(3-2q^2)}{\displaystyle 8q^3}\Erf(q), 
\\
Q_4(q)&=&-\frac{\displaystyle 5(21+2q^2)}{\displaystyle 32q^4}e^{-q^2}
+\frac{\displaystyle 3\sqrt{\pi}(35-20q^2+4q^4)}{\displaystyle 64q^5}\Erf(q), 
\\
Q_6(q)&=&-\frac{\displaystyle 21(165+20q^2+4q^4)}{\displaystyle 128q^6}e^{-q^2}
+\frac{\displaystyle 5\sqrt{\pi}(693-378q^2+84q^4-8q^6)}{\displaystyle 256q^7}\Erf(q).
\label{Q6express}
\end{eqnarray}
The error function has the asymptotic form $\Erf(q)\rightarrow 1$ for
$q\gg1$.  Therefore, $Q_\ell(q)$ is in proportion to
$(-1)^{\ell/2}q^{-1}$ in the limit $q\gg1$, which explains the
asymptotic behavior of the multipole power spectrum at the large wave
numbers. The central-satellite contribution is dominant in the 1-halo
term of LRG samples and thus $kP_\ell^{\rm 1h}\sim 2kQ_\ell(q)f_{\rm
  sat}/\bar{n}$ where $f_{\rm sat}$ is the satellite fraction.  When
we use the values of $\bar{\sigma}_{v,{\rm off}}= 663$km/s,
$aH(z)=88h$km/s at $z=0.32$, $f_{\rm sat}=0.07$ and $\bar{n}\simeq
10^{-4}({\rm Mpc/h})^{-3}$ for the LRG sample, the large-scale limit
of $kP_\ell^{\rm 1h}$ goes to $230$ for $\ell=0$, $-120$ for $\ell=2$,
$85$ for $\ell=4$, and $-74$ for $\ell=6$. Figure~\ref{fig:model_all}
shows that 1-halo term contribution approaches these values roughly.

Figure~\ref{fig:model_sat} compares the halo model predictions of the
multipole power spectra for the ``NBLRG'' sample and the
observed spectra. As 40\% of BLRGs in multiple LRG systems are
satellites (the number of multiple LRG systems ${\cal N}_{\rm mul}$ is
4157), the same number of central galaxies are mixed in the NBLRG
sample.  Here we consider that $35$\% (=$0.4{\cal N}_{\rm mul}/{\cal
  N}_{\rm sat}$) of the NBLRG sample are central galaxies and the rest
of them are satellites.  Based on this assumption we write the
one-halo and two-halo terms of the power spectrum of the NBLRG sample
as
\begin{eqnarray}
&&P^{\rm NBLRG}(k,\mu)=P^{\rm 1h,NBLRG}(k,\mu)+P^{\rm 2h,NBLRG}(k,\mu), \\
&&P^{\rm 1h,NBLRG}(k,\mu)=\frac{1}{\bar{n}_{\rm sat}^2}\int\!dM~\frac{dn}{dM}
\langle N_{\rm cen}\rangle \langle N_{\rm sat}\rangle^2~~~~~~~~~~~~~~~~~~~~~ \nonumber \\
&&~~~~~~~~~~~~~~~~~\times\left[2q_{\rm cen}^{\rm (sat)}(1-q_{\rm cen}^{\rm (sat)})\tilde{p}_{\rm cs}(k,\mu; M)
+(1-q_{\rm cen}^{\rm (sat)})^2\tilde{p}_{\rm ss}(k,\mu; M)\right], 
\\
&&P^{\rm 2h,NBLRG}(k,\mu)=\biggl[\frac{1}{\bar{n}_{\rm sat}}\int\!dM~
\frac{dn}{dM}(b(M)+f\mu^2)
\langle N_{\rm cen}\rangle\langle N_{\rm sat}\rangle\nonumber \\
&&~~~~~~~~~~~~~~~~~\times (q_{\rm cen}^{\rm (sat)}+
(1-q_{\rm cen}^{\rm (sat)})\tilde{p}_{\rm cs}(k,\mu; M))
\tilde{u}_{\rm vol}(k;M)\biggr]^2P^{\rm NL}_{\rm m}(k), 
\label{eq:pksat}
\end{eqnarray}
where $q_{\rm cen}^{\rm (sat)}$ is the fraction of central galaxies in
the NBLRG sample. Here we set $q_{\rm cen}^{\rm (sat)}={\rm
  Min}(0.35,1/\langle N_{\rm sat} \rangle)$ so that $q_{\rm cen}^{\rm
  (sat)}\langle N_{\rm sat}\rangle$ does not exceed unity.  The
one-halo and two-halo terms for NBLRGs are plotted with the dotted
curve and the dashed curve, respectively. 
The halo model explains the observed multipole spectral very well.
One-halo term (dotted curve) is a dominant contribution to the
multipole spectra and reaches $kP_\ell^{\rm NBLRG}(k)\sim {\cal
  O}(10^3)$.

\begin{figure*}[b]
\begin{center}
\includegraphics[scale=.35]{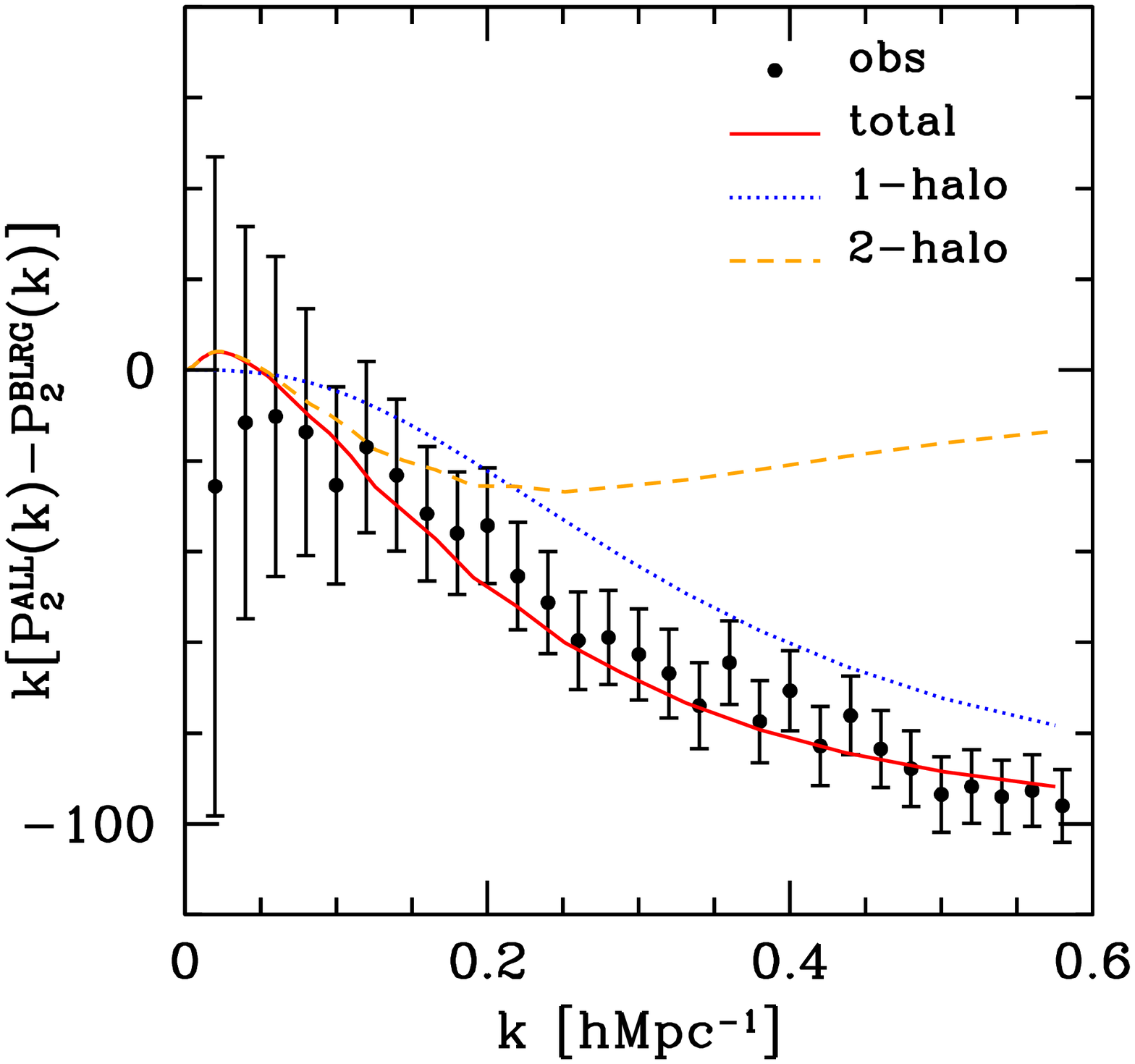}
\includegraphics[scale=.35]{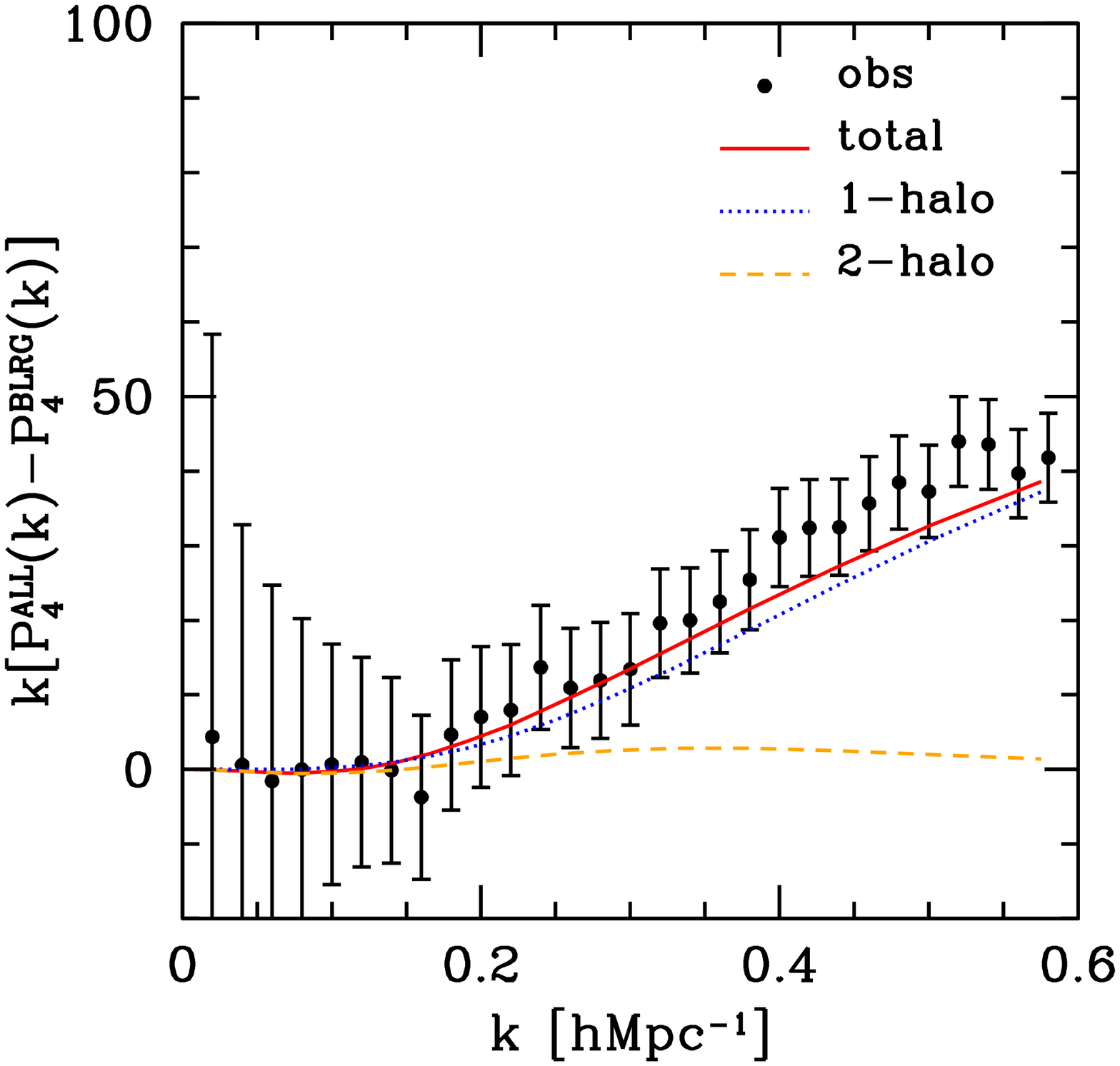}
\caption{Differences of multipole power spectra $P_2$ (left) and $P_4$
  (right) between All and BLRG samples. Sold curves show the halo
  model prediction, which mainly comes from the one-halo term (dotted
  curve) compared to the two-halo term (dashed curve).
\label{fig:model_sub}
}
\end{center}
\end{figure*}

The brightest LRG power spectrum does not have one-halo term because
each halo contains one LRG at most. Then, the BLRG power spectrum is
written only with the two-halo term. This can be clearly seen in
that the BLRG multipole power spectra with $\ell=4$ and $6$
are almost zero in Figure \ref{fig:model_blrg}, 
which indicates that the 1-halo term from satellite
galaxies becomes significantly smaller by removing NBLRGs. This
means that the halo reconstruction method we use well succeeds in
removing the one-halo term. This is important for the study of
precision cosmology because the uncertainty of the satellite HOD and
its FoG effect becomes significantly small. In a strict sense,
however, $P_4(k)$ has slightly positive signature compared to the halo
model predictions. This may come from that the reconstruction method
is incomplete and some of satellite LRGs are included in the BLRG
sample. Multipole spectra such as $P_4(k)$ dominated by the 1-halo 
term is useful for estimating the residual 1-halo term effect.

However, several observations indicate that some fraction of BLRGs are
satellite or off-centered galaxies (e.g., \cite{Skibba2011}), which
causes the FoG effect \cite{Hikage2011}.  When the fraction that BLRGs
locate on the mass center of their host halos is $q_{\rm cen}^{\rm
  (BLRG)}$, the BLRG power spectrum is given
\begin{eqnarray}
&&P^{\rm BLRG}(k,\mu)=\biggl[\frac{1}{\bar{n}}\int\!dM~\frac{dn}{dM}(b(M)+f\mu^2)
\langle N_{\rm cen}\rangle~~~~~~~~~~~~ \nonumber \\
&&~~~~~~~~~~~~~~~~~~~~\times(q_{\rm cen}^{\rm (BLRG)}+(1-q_{\rm cen}^{\rm (BLRG)})
\tilde{p}_{\rm cs}(k,\mu; M))\tilde{u}_{\rm vol}
\biggr]^2P^{\rm NL}_{\rm m}(k).
\label{eq:pk_2h_BLRG}
\end{eqnarray}
As shown in \cite{Hikage2012}, the lensing and cross-correlation
measurements indicate 20\% fraction of BLRGs are off-centered. 
Figure~\ref{fig:model_blrg} shows the model predictions of the BLRG power
spectra with $q_{\rm cen}^{\rm (BLRG)}=1$ (all of BLRGs are centrals)
and $q_{\rm cen}^{\rm (BLRG)}=0.8$ (20\% of BLRGs are satellite). The
figure shows that their agreement is better at high $k$ when the satellite FoG
effect is included, which indicates that even BLRG sample may have
significant FoG effect on the multipole power spectra. Note that the
result may change if we take into account the nonlinearity in the
galaxy biasing. The linear Kaiser formula is the simplest model, then
more careful analysis will be necessary using the sophisticated
perturbation theories as well as numerical simulations
(\cite{Scoccimarro2004,Matsubara2008,Taruya2010,Nishimichi2011}),
though such analysis is beyond the scope of the present paper.

Figure \ref{fig:model_sub} shows the differences of multipole power
spectra $P_2$ and $P_4$ between ALL and BLRG samples. The curves in
each panel are the theoretical prediction of our model using the
satellite HOD parameters of the NBLRG samples. The theoretical curves
much better fit the observational results, compared with those in
Figure \ref{fig:model_all}.  This agreement indicates the
contamination of the FoG effect of the off-centered velocities in the
BLRGs.

\section{Constraints on the growth rate and the properties of satellite galaxies}
\label{sec:constraint}
In this section, we consider a constraint by comparing
the observed multipole power spectrum of the LRG samples 
and our theoretical model including the one-halo term. 
This demonstrates how the one-halo term influences a cosmological 
constraint. We define $\chi^2$ by 
\begin{eqnarray}
 \chi^2=\sum_{\ell=0,2,4,6}\sum_{i}{[P^{\rm obs.}_\ell(k_i)-P^{\rm model}_\ell(k_i)]^2\over 
 [\Delta P_\ell(k_i)]^2},
\label{chi222}
\end{eqnarray}
where $P^{\rm obs.}_\ell(k_i)$ and $\Delta P_\ell(k_i)$ are the observed 
power spectrum and the error, respectively, and 
$P^{\rm model}_\ell(k_i)$ is the theoretical model, described in the below.

Based on the halo model developed in previous section, we fit the
observed power spectra with the following form of the power spectra
averaged over halo mass
\begin{eqnarray}
&&P^{\rm model}(k,\mu)=P^{\rm 1h,model}(k,\mu)+P^{\rm 2h,model}(k,\mu) \\
&&P^{\rm 1h,model}(k,\mu)=
\frac{2f_{\rm sat}}{\bar{n}}{\cal D}\left(\frac{k\mu\bar\sigma_{v,{\rm off}}}{aH}\right), \\
&&P^{\rm 2h,model}(k,\mu)=
\left\{\left(\bar{b}(k)+f\mu^2\right)
\left[\left(1-f_{\rm sat}\right)+f_{\rm sat}
{\cal D}\left(\frac{k\mu\bar\sigma_{v,{\rm off}}}{aH}\right)\right]\right\}^2P_{\rm m}^{\rm NL}(k), 
\label{eq:fit}
\end{eqnarray}
where $\bar{b}(k)$ is the averaged bias of LRGs and linearly fitted as
$b_0+b_1k$. Here the growth rate is $f=\Omega_m(z)^\gamma$, assuming
the $\Lambda$CDM model as background universe, and the other
cosmological parameters are fixed as $n_s=0.96$, $\Omega_m=0.28$,
$\Omega_b=0.044$, $\sigma_8=0.8$. We consider the FoG of satellite
LRGs and parametrize it with the satellite fraction $f_{\rm sat}$ and
the averaged velocity dispersion $\bar\sigma_{v,{\rm off}}$. We use a
Lorentzian form of FoG damping function of (\ref{eq:calD}), which well
approximates the observed satellite velocity distribution as shown in
Figure \ref{fig:veldif}. In the limit of small $k$, the FoG damping
function ${\cal D}(x)$, eq.~(\ref{eq:calD}), becomes $1-x^2/2$. In
this lowest-order approximation, $\widetilde\sigma_v^2$ in equation
(\ref{eq:pkfit}) corresponds to $2f_{\rm sat}(\bar\sigma_{v,{\rm
    off}}H_0/aH(z))^2$ in equation (\ref{eq:fit}). For the 1-halo
term, we only take the dominant contribution of the central-satellite
pairs into account, and neglect that from the satellite-satellite
pairs. We do not introduce additional parameter of central fraction
(i.e., $q_{\rm cen}$), for simplicity, while it is still controversial
issue whether BLRGs are off-centered or not.  Instead, we leave
$f_{\rm sat}$ as a free parameter because the observed multipole power
spectrum, $P_4(k)$ in Figure \ref{fig:model_blrg}, systematically
deviates from zero even at small $k$, which suggests the residual
1-halo terms.  In summary, the number of the fitting parameters is $5$
in total: $b_0, b_1, \gamma, f_{\rm sat}$, and $\bar\sigma_{v,{\rm
    off}}$.
\begin{table*}[t]
\begin{center}
\begin{tabular}{lcccccc}
\hline
\hline
~ & method  & Sample & $\gamma$ & $100f_{\rm sat}$ & $\bar\sigma_{v,{\rm off}}$[km/s] \\
\hline
~ &$k<0.2h$/Mpc for all $P_l$
     & All    & $0.78\pm 0.10$ & $ 20\pm  21$ & $ 590\pm 300$ \\
(I) &w/o 1-halo term
  &    BLRG   & $0.69\pm 0.07$ & $ 33\pm  30$ & $ 210\pm 250$ \\
  &   & Single & $0.62\pm 0.06$ & $ 33\pm  30$ & $ 200\pm 220$ \\
\hline
~ &$k<0.2h$/Mpc for all $P_l$
     & All    & $0.72\pm 0.10$ & $4.5\pm 1.6$ & $ 910\pm 180$ \\
(II) & with 1-halo term
   &   BLRG   & $0.64\pm 0.08$ & $1.2\pm 1.3$ & $ 700\pm 320$ \\
   &  & Single & $0.60\pm 0.07$ & $1.0\pm 1.5$ & $ 590\pm 340$ \\
\hline
~& $k<0.2h$/Mpc for $P_0, P_2$
     & All    & $0.54\pm 0.04$ & $8.5\pm 0.4$ & $780\pm   50$ \\
(III) &$k<0.6h$/Mpc for $P_4, P_6$ 
     & BLRG   & $0.57\pm 0.04$ & $2.4\pm 0.4$ & $850\pm  150$ \\
 & with 1-halo term
     & Single & $0.56\pm 0.04$ & $1.8\pm 0.5$ & $830\pm  190$ \\

\hline
\end{tabular}
\caption{Constraints on the index of growth rate $\gamma$, satellite
  fraction $f_{\rm sat}$ and averaged velocity dispersion
  $\bar\sigma_{v,{\rm off}}$ from the fitting of the multipole power
  spectra $P_l(k)$ with $(l=0,2,4,6)$ for All, BLRG and Single LRG
  samples.  In the fitting, we compare the three methods with and
  without 1-halo term in the modeling and adopting the different range
  of wavenumbers: (I) fitting all $P_l(k)$ in the range of
  $k<0.2h/$Mpc without 1-halo term (top); (II) fitting all $P_l(k)$ in
  the range of $k<0.2h/$Mpc with 1-halo term (middle); (III) fitting
  $P_0(k)$ and $P_2(k)$ in the range of $k<0.2h/$Mpc while $P_4(k)$
  and $P_6(k)$ in the range of $k<0.6h/$Mpc with 1-halo term (bottom).
\label{tab:limit}
}
\end{center}
\end{table*}

\begin{figure*}[b]
\begin{center}
\includegraphics[width=15.5cm]{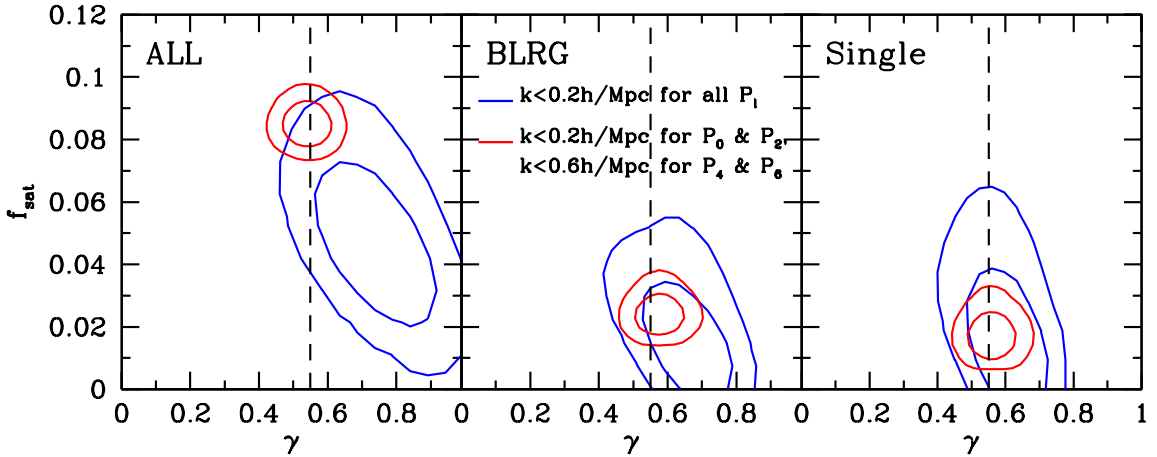}
\caption{Joint constraints on $\gamma$ and satellite fraction $f_{\rm
  sat}$ from the fitting of the multipole power spectra $P_l(k)$ 
  with $(l=0,2,4,6)$ for All, BLRG, and Single samples.
  In each panel, the (blue) large curves are the $1\sigma$ and $2\sigma$
  contours with the data of the range $k<0.2h/$Mpc, while
  the (red) small circles are the same but with the data 
  of the range $k<0.6h/$Mpc for $P_4(k)$ and $P_6(k)$.
  The vertical dashed line shows $\gamma=0.55$, 
  the prediction of the general relativity.
\label{fig:cont}}
\end{center}
\end{figure*}
\begin{figure*}
\begin{center}
\includegraphics[width=15.5cm]{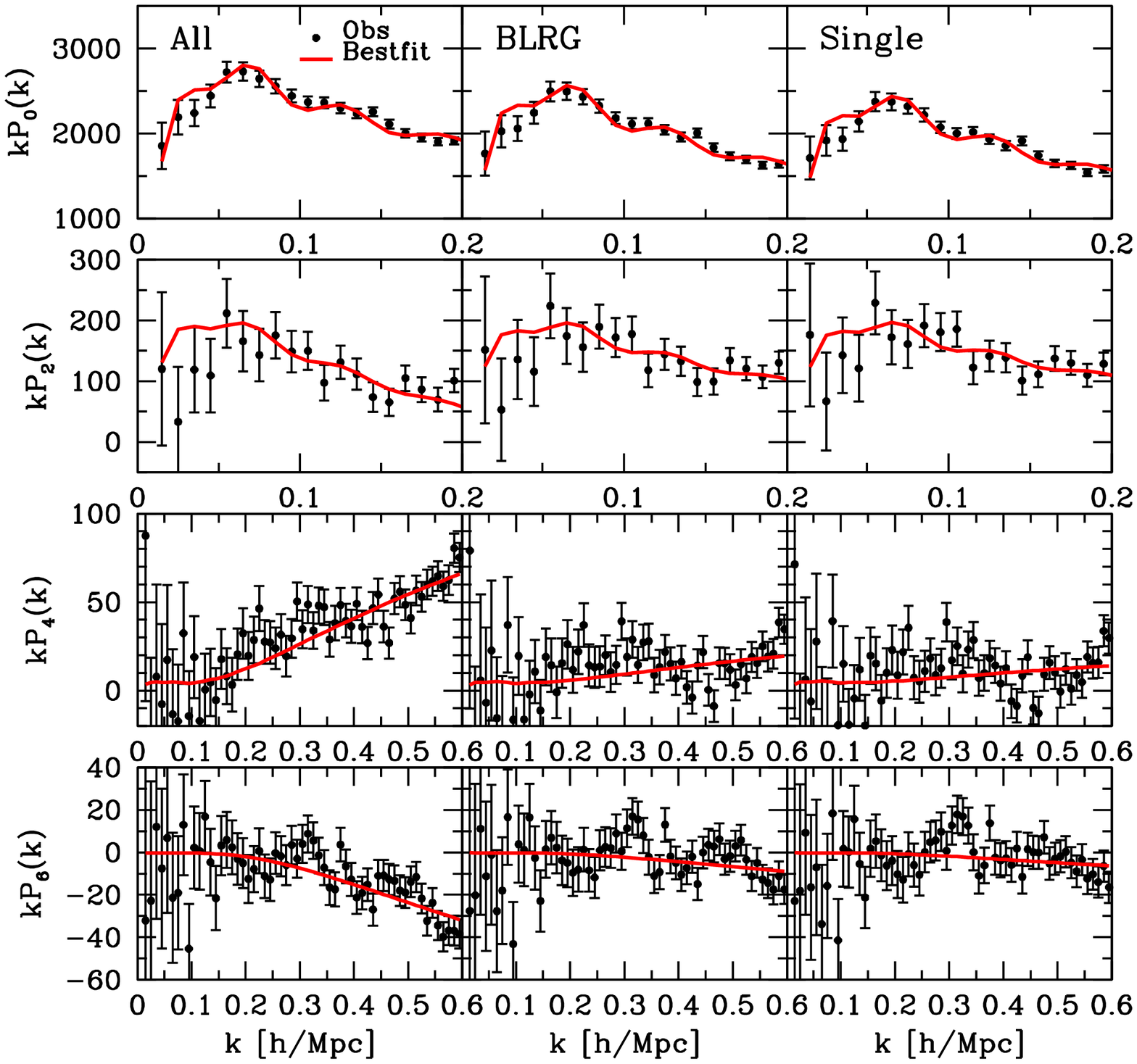}
\caption{Comparison of $P_0(k)$, $P_2(k)$, $P_4(k)$ and $P_6(k)$ for
  All (left panel), BLRG (center panel), and Single (right panel)
  samples and the models with the best-fit parameters. The maximum
  value of $k$ is $0.2h$/Mpc for $P_0(k)$ and $P_2(k)$, while
  $0.6h$/Mpc for $P_4(k)$ and $P_6(k)$.
\label{fig:pkbest}}
\end{center}
\end{figure*}

First, let's see how adding the 1-halo term in the theoretical model
changes the fitting results.  Table \ref{tab:limit} compares the
constraints on the parameters of $\gamma, f_{\rm sat},
\bar\sigma_{v,{\rm off}}$ without 1-halo term and those with 1-halo
term for All, BLRG, and Single LRG samples, respectively.  The fitting
range is up to $k=0.2h/$Mpc for all of $P_l (l=0,2,4,6)$, and the bias
parameters are marginalized over. In the fitting (I) without 1-halo
term, the value of $\gamma$ for All sample is overestimated (or growth
rate $f$ is underestimated) compared to that of the BLRG or Single LRG
sample, which is also shown in Figure \ref{fig:impact}: the difference
of best-fit values of $\gamma$ between All and Single is 0.16 and that
between BLRG and Single is 0.07. The deviation is mildly alleviated by
including the 1-halo term in the fitting (II): 0.12 between ALL and
Single and 0.04 between BLRG and Single. However, the 1-halo term
effect is highly degenerated with the growth rate.

Next we add the information of the small-scale measurements of
$P_4(k)$ and $P_6(k)$ of the range of wavenumbers up to $k=0.6h$/Mpc
in the fitting. As shown in the previous section, $P_4(k)$ and
$P_6(k)$ at large $k$ is dominated by the FoG effect of 1-halo term,
then the information can be used to calibrate the uncertainty of the
satellite FoG. We find that the information of $P_4(k)$ and $P_6(k)$
on small scales (at large $k$) significantly improves the error of
satellite fraction by a factor $3\sim 4$ and the error of $\gamma$ by
a factor 2. Here $P_4(k)$ plays an important role, especially.  Our
constraints on $\gamma$ from the 3 different LRG samples becomes
consistent with each other by including the higher multipole spectra
at large $k$. This indicates that our fitting formula including 1-halo
term well describes the behavior of three different LRG
samples. Figure \ref{fig:cont} shows the contour of the joint
constraints on $\gamma$ and $f_{\rm sat}$ when the small-scale
information of $P_4(k)$ and $P_6(k)$ is included (red) and not
included (blue).  It is clearly seen that the measurements of $P_4(k)$
and $P_6(k)$ on small-scales break the degeneracy between $\gamma$ and
$f_{\rm sat}$ and improves their errors dramatically.  Figure
\ref{fig:pkbest} compares the observations (black filled circles with
error bars) of the multipole power spectra and the corresponding
best-fitted curve (red solid curves). Our model well describes the
observations of the three samples including $P_4(k)$ and $P_6(k)$ at
large $k$.  Actually the satellite fraction for BLRG and Single LRG
samples significantly decreases, as described in Table
\ref{tab:limit}.  However, it still remains $\sim$ 2\% fraction of
central-satellite pair, accordingly the satellite fraction for All
sample becomes $\sim 7$ \%, which is higher than the expected value
including the fiber collision $\sim 5\%$.  This may indicate that the
halo reconstruction is incomplete and some of satellite galaxies are
still included in BLRG and Single LRG samples. Multipole power spectra
such as $P_4(k)$ and $P_6(k)$ are a good indicator for the residual
1-halo term and may be useful for finding a better grouping method.
Our constraint on the velocity dispersion of $\bar\sigma_{v,{\rm off}}$
is $\sim 800$km/s, which is roughly consistent with the observed
pairwise velocity dispersion between BLRG and NBLRGs, that is $653$km/s
as shown in Figure \ref{fig:veldif}. 

Our method using the measurements of higher multipole spectrum
$P_4(k)$ provides a promising way to calibrate the satellite FoG
effect and improve the error of the growth rate measurement.  The
measurements of the satellite fraction and the velocity dispersion can
be translated to the constraints on the satellite HOD and/or the
velocity bias between LRGs and halos.  However, our theoretical model
is still very simple and uses various approximations such as Kaiser
approximation. In order to obtain more robust estimates on the growth
rate and satellite properties, we need more precise theoretical models
of halo clustering and velocity probability distribution function by
comparing with simulated mock samples. Such detailed analysis is
beyond the scope of this paper and is left as future work.

\section{Forecast on multipole power spectra for H$\alpha$ emitters} 
\label{sec:forecast}

Main targets of high-redshift ($z=1\sim 2$) galaxy surveys, planed in
such as Subaru/PFS \cite{PFS2012} and Euclid \cite{Euclid2012}, are
H$\alpha$ emitters (HAE). In this section, we perform Fisher analysis
to estimate the impact of satellite galaxies for such future surveys
targeting HAEs.

\subsection{HOD of H$\alpha$ emitters}
\begin{figure*}
\begin{center}
\includegraphics[width=16cm]{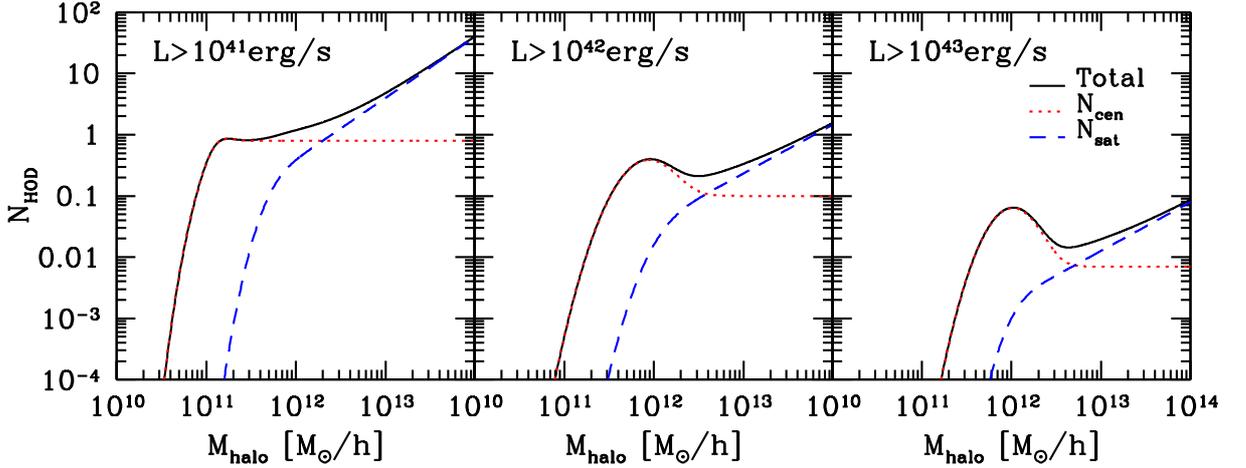}
\caption{HOD for H$_\alpha$ emitters based on \cite{Geach2012}.
Here we set no fiber collision effect $f_{\rm col}=1$.
\label{fig:hod_halpha}}
\end{center}
\end{figure*}

The relation of HAEs to halos are less known observationally, and will
be more complicated than that of LRGs.  We use the following form of
HOD based on the sample of 370 HAEs at z=2.23 detected in Hi-Z
Emission Line Surveys (HiZELs) \cite{Geach2012}
\begin{eqnarray}
&&\langle N_{\rm cen}^{H\alpha}\rangle =F_b(1-F_a)\exp\left[-\frac{(\log_{10}(M)-\log_{10}(M_{\rm min,c}))^2}{2\sigma_{\log M,{\rm c}}^2}\right]~~ \nonumber\\
&&~~~~~~~~~~+F_a\left[1+{\rm erf}\left(\frac{\log_{10}(M)-\log_{10}
(M_{\rm min,c})}{\sigma_{\log M,{\rm c}}}\right)\right], \\
&&\langle N_{\rm sat}^{H\alpha}\rangle =f_{\rm col}F_s\left[1+{\rm erf}\left(\frac{\log_{10}(M)-\log_{10}(M_{\rm min,s})}{\sigma_{\log M,{\rm s}}}\right)\right]\left(\frac{M}{M_{\rm min,s}}\right)^{\alpha}.
\label{eq:HOD_HAE}
\end{eqnarray}
Here the central HAE distribution is described with Gaussian and
smoothed step-like components with their amplitudes determined by the
normalization factors $F_a$ and $F_b$. The typical mass and the
dispersion are parametrized with $M_{\rm min,c}$ and $\sigma_{\log
  M,{\rm c}}$, respectively. Satellite HOD is described with a
smoothed step-like component multiplied by power-law with scaling of
$\alpha$, the typical satellite mass $M_{\rm min,s}$ and the amplitude
$F_s$.  The values of HOD parameters for different luminosity samples
are listed in Table \label{tab:hod_HAE}. 
The plots of HODs are shown in Figure~\ref{fig:hod_halpha}.

\begin{table}[b]
\caption{HOD parameters for HAEs\label{tab:hod_HAE}}
\begin{center}
\begin{tabular}{lrrr}
\hline
Luminosity & $L>10^{41}$erg/s & $L>10^{42}$erg/s & $L>10^{43}$erg/s \\
\hline
$F_a$ & 0.4 & 0.05 & 0.0035 \\
$F_b$ & 0.33 & 0.35 & 0.06 \\
$F_s$ & 0.1 & 0.02 & 0.001 \\
$M_{\rm min,c}(10^{12}M_\odot/h)$ & 0.12 & 0.8 & 1 \\
$\sigma_{\log M, {\rm c}}$ & 0.14 & 0.25 & 0.22    \\
$M_{\rm min,s}(10^{12}M_\odot/h)$ & 0.5 & 1.1 & 1  \\
$\sigma_{\log M, {\rm s}}$ & 0.24 & 0.32 & 0.24 \\
$\alpha$ & 1 & 0.8 & 0.8 \\
\hline 
\end{tabular}
\end{center}
\end{table}

\begin{figure*}
\begin{center}
\includegraphics[scale=.35]{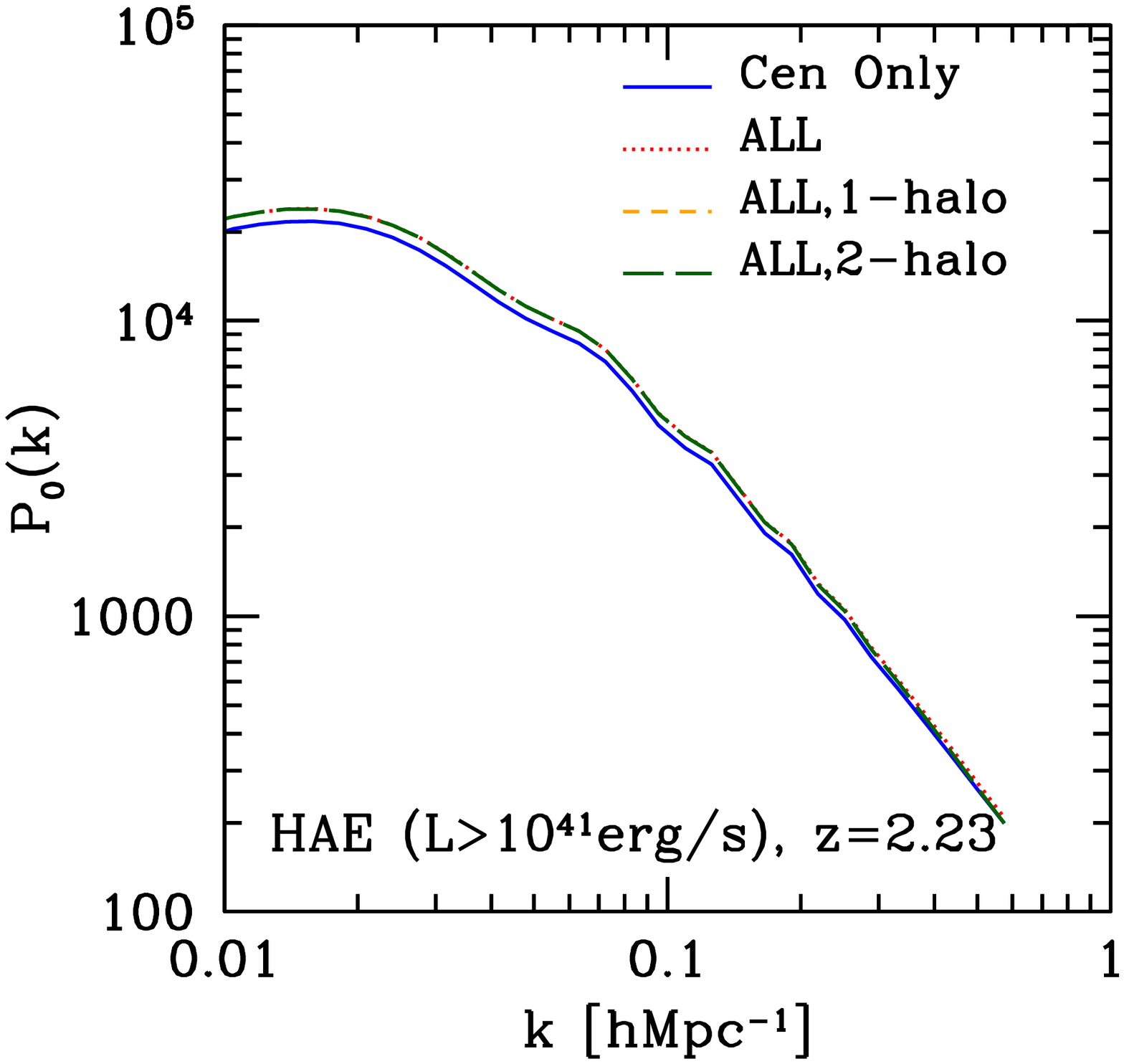}
\includegraphics[scale=.35]{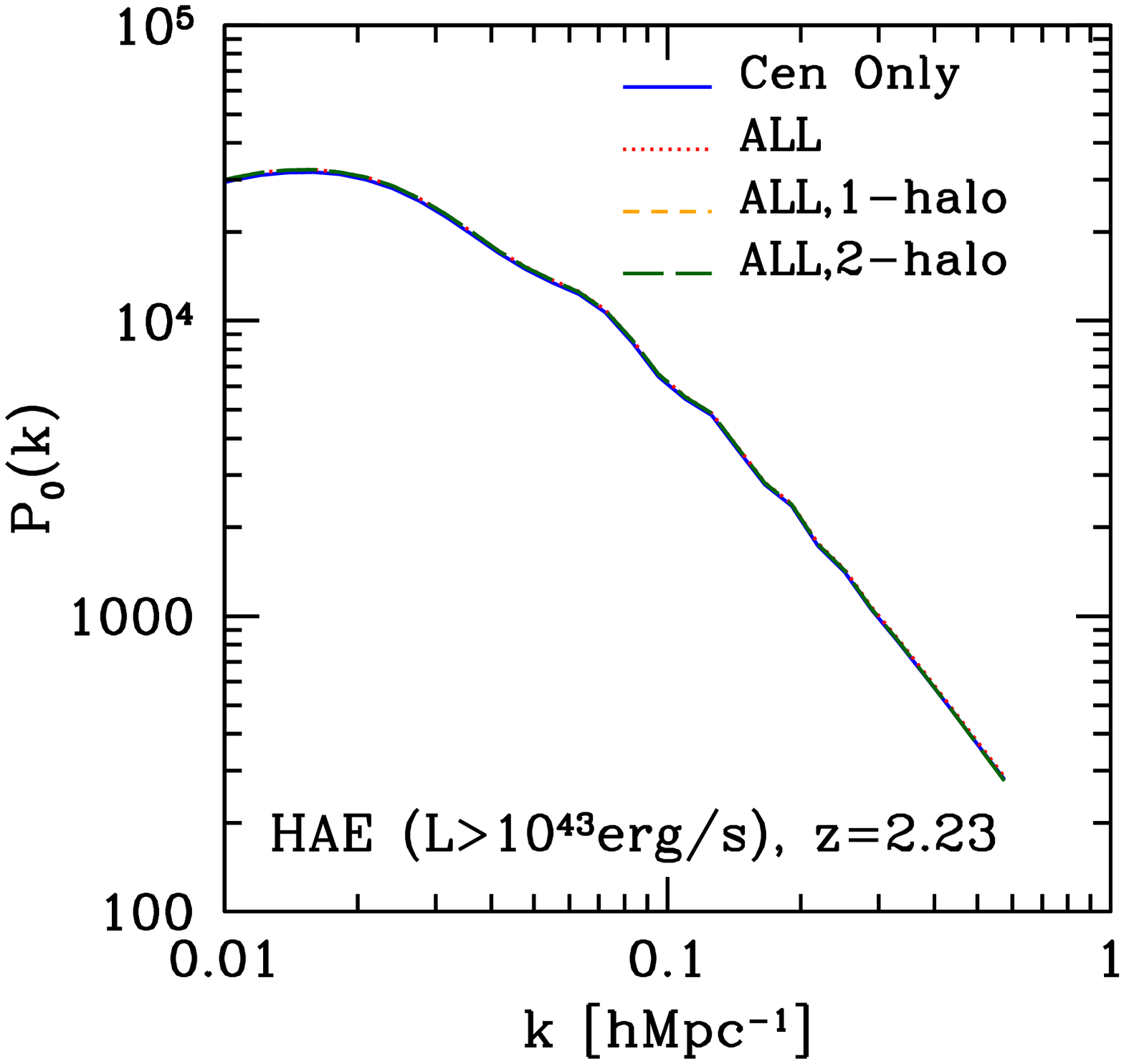}
\includegraphics[scale=.35]{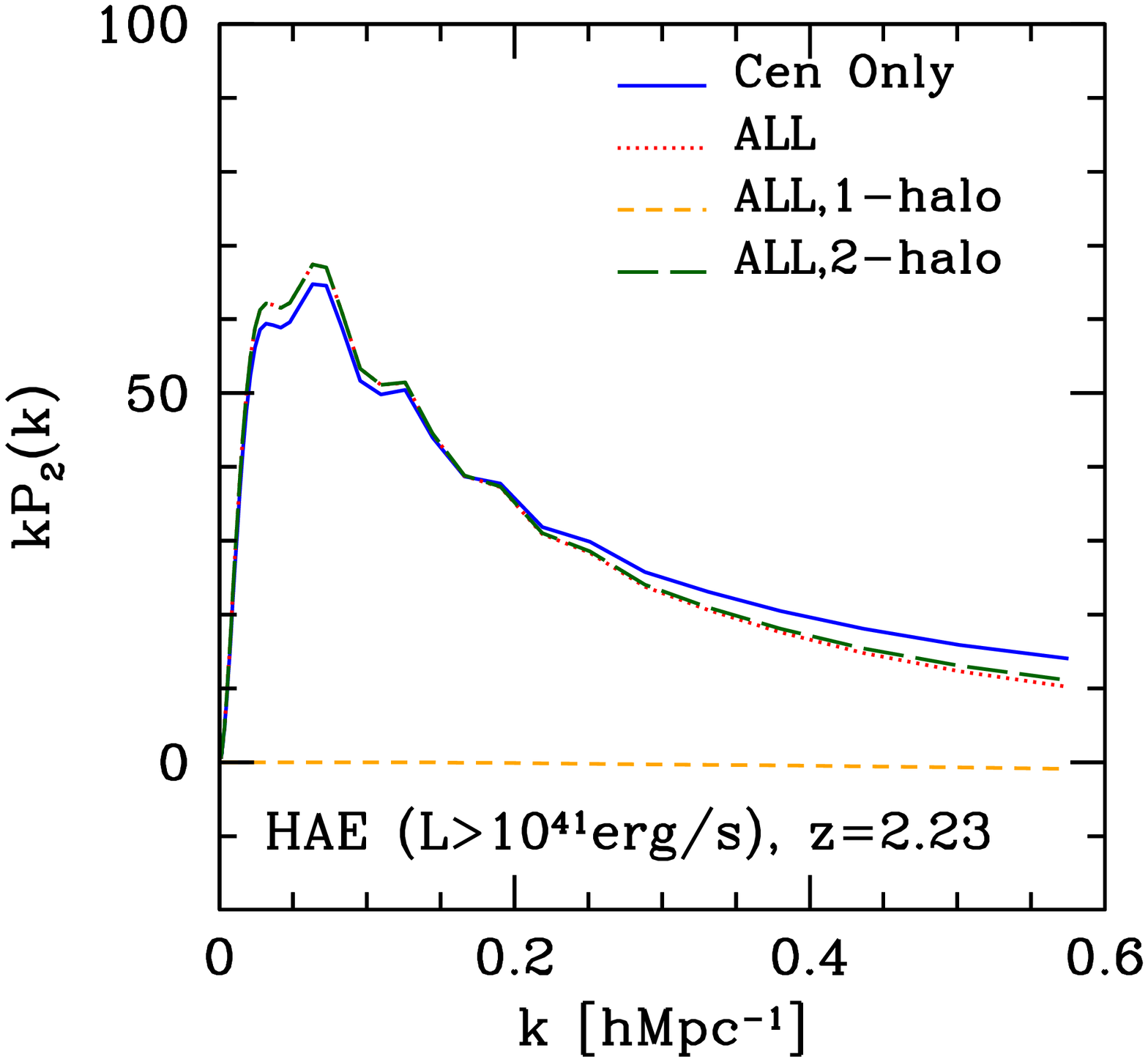}
\includegraphics[scale=.35]{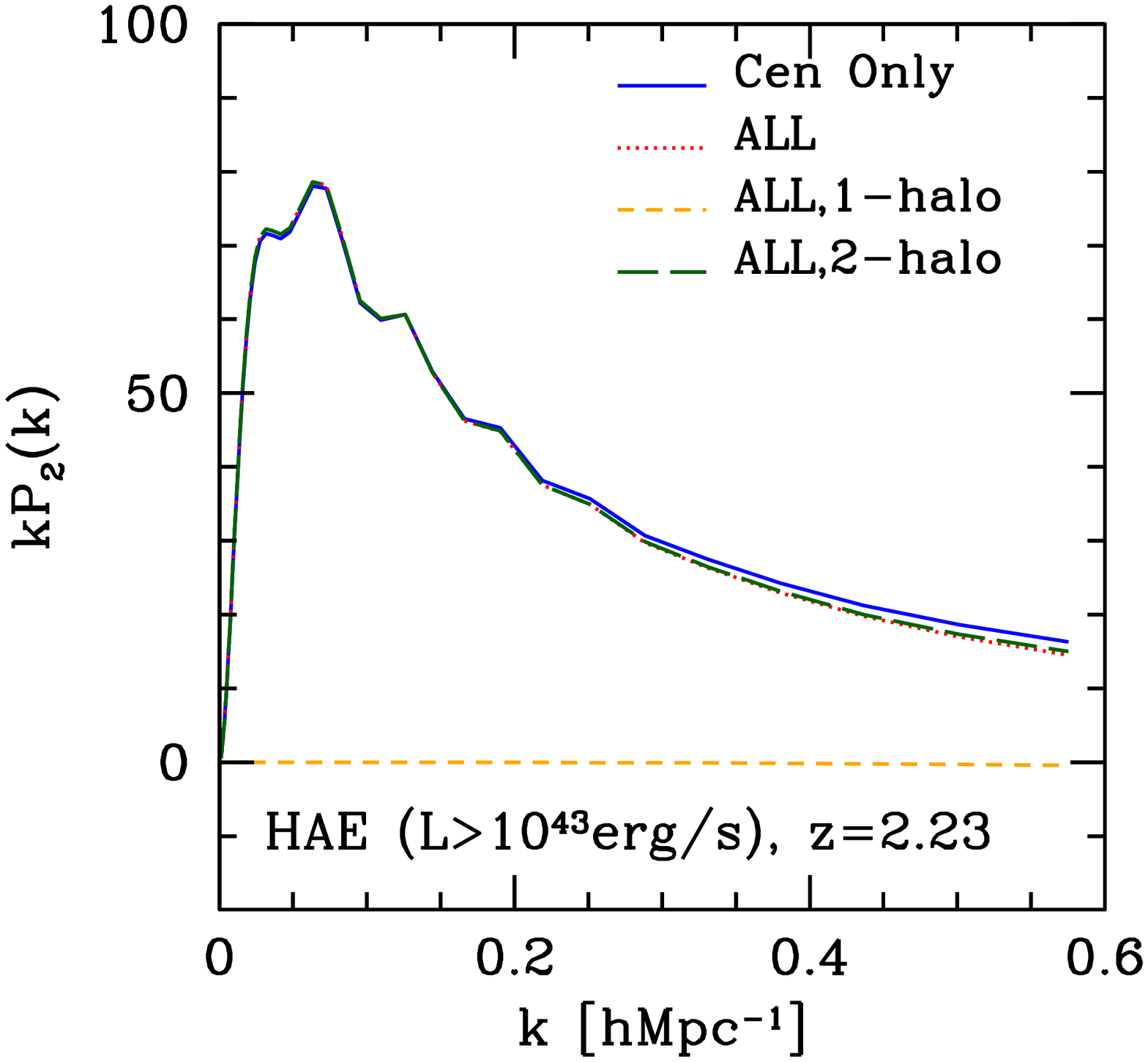}
\includegraphics[scale=.35]{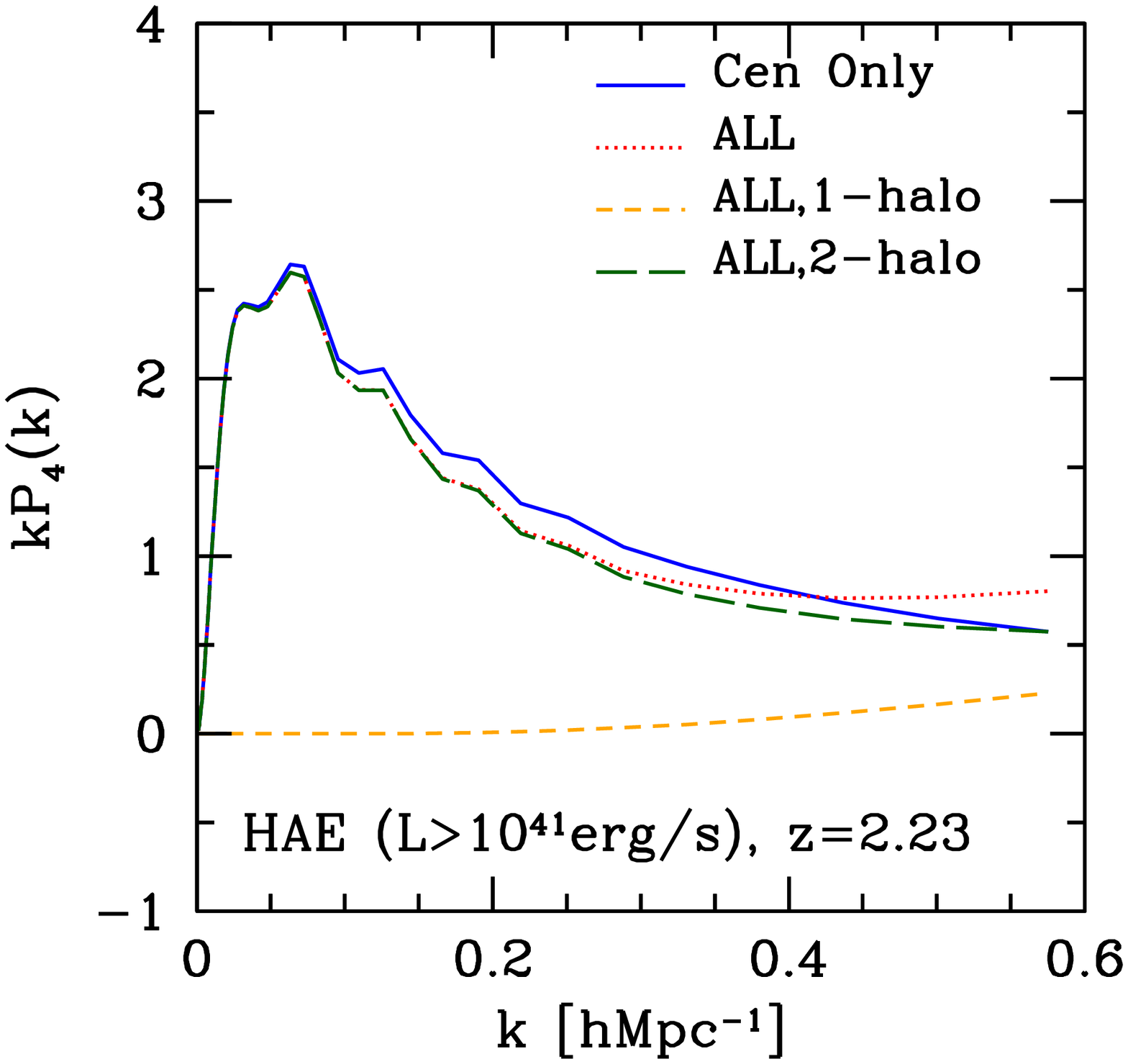}
\includegraphics[scale=.35]{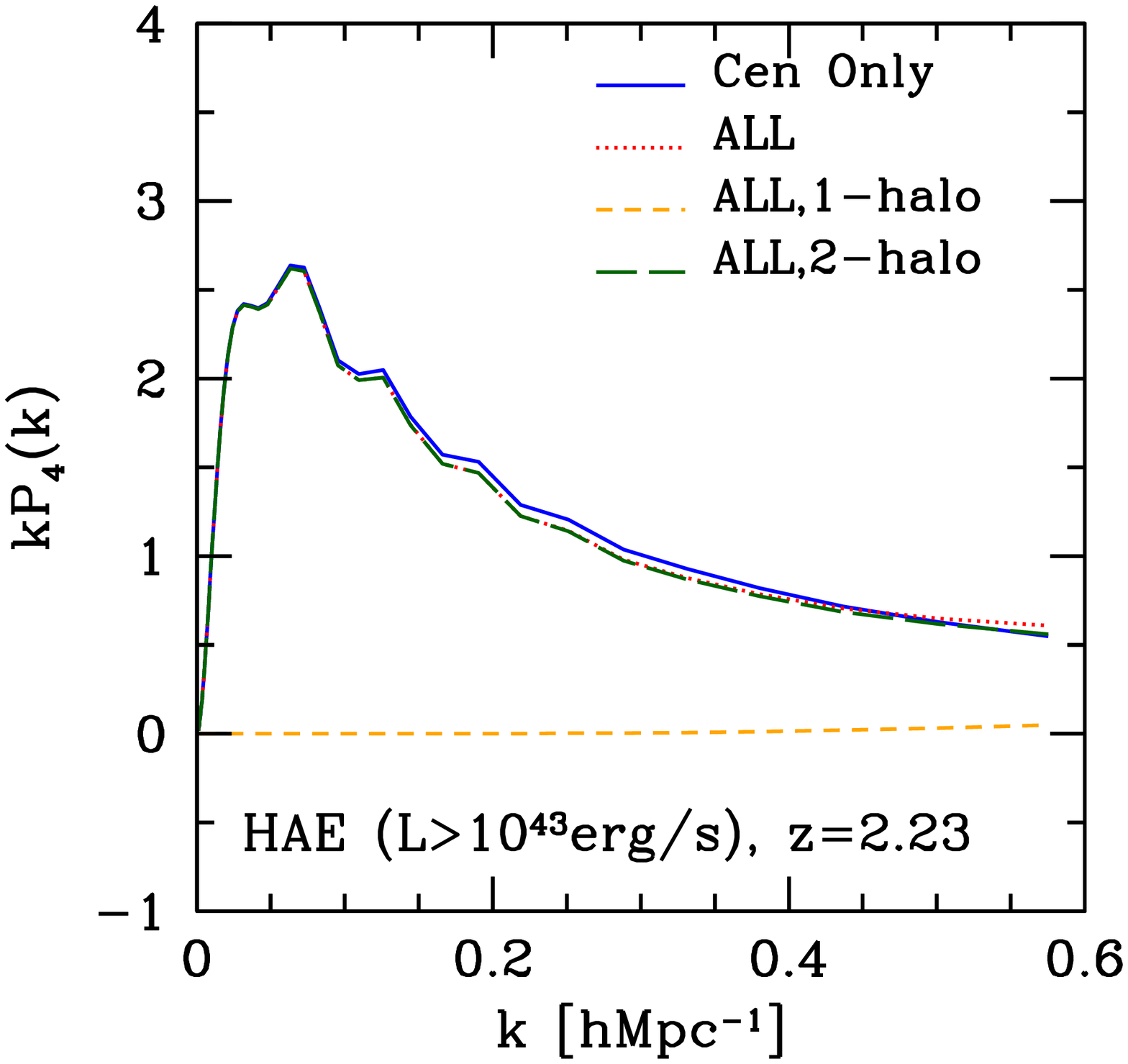}
\caption{Multipole power spectra $P_0$ (top), $P_2$ (middle), and $P_4$
  (bottom) for H$_\alpha$ emitters with $L>10^{43}$erg/s (left panels)
  and $L>10^{41}$erg/s (right panels) at $z=2.23$.
\label{fig:model_Ha2}
}
\end{center}
\end{figure*}

Figure~\ref{fig:model_Ha2} shows the comparison of multipole power
spectra for central HAEs with those for all HAEs including satellites
at $z=2.23$. The FoG effect on the HAE power spectrum from satellite
in a halo is much smaller than that on LRGs power spectrum because the
typical halo mass of HAEs is much smaller than that of LRGs. In our
halo model, averaged virial velocity of halos hosting HAEs is 170km/s,
while those hosting LRGs are 660km/s. For the faint HAE sample, the
contamination of the satellite changes the higher multipole spectrum
at a few percent or 10 percent level depending on the wave number,
while the effect becomes smaller for the luminous HAE sample because
the satellite fraction decreases. FoG effect for HAEs are expected to
be much smaller than LRGs, however, upcoming galaxy surveys are
expected to measure the growth rate measurement at the percent-level
accuracy and thus it is still important to estimate the systematic
errors of the FoG effect.

\begin{figure*}[t]
\begin{center}
\includegraphics[scale=.45]{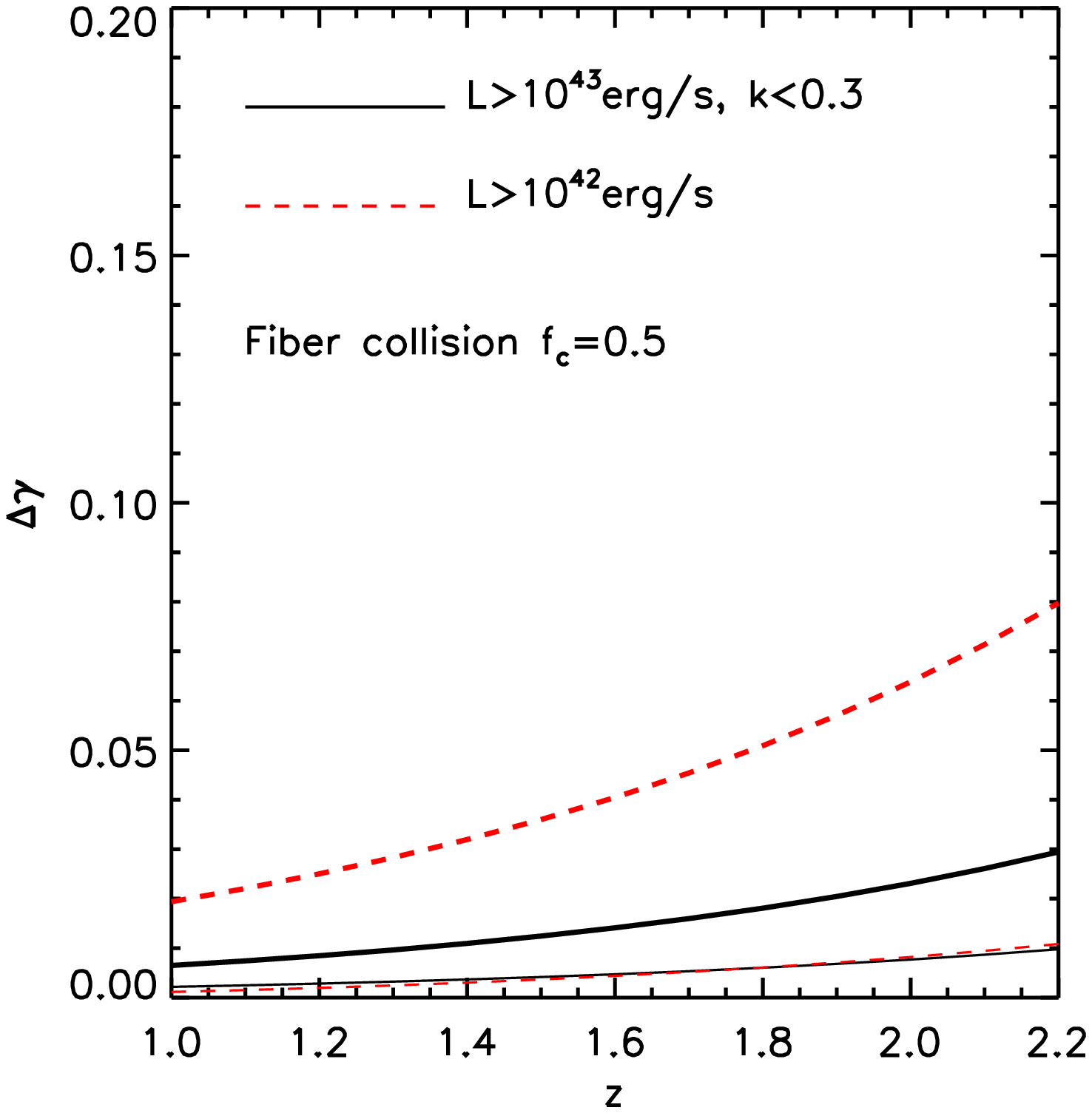}
\includegraphics[scale=.45]{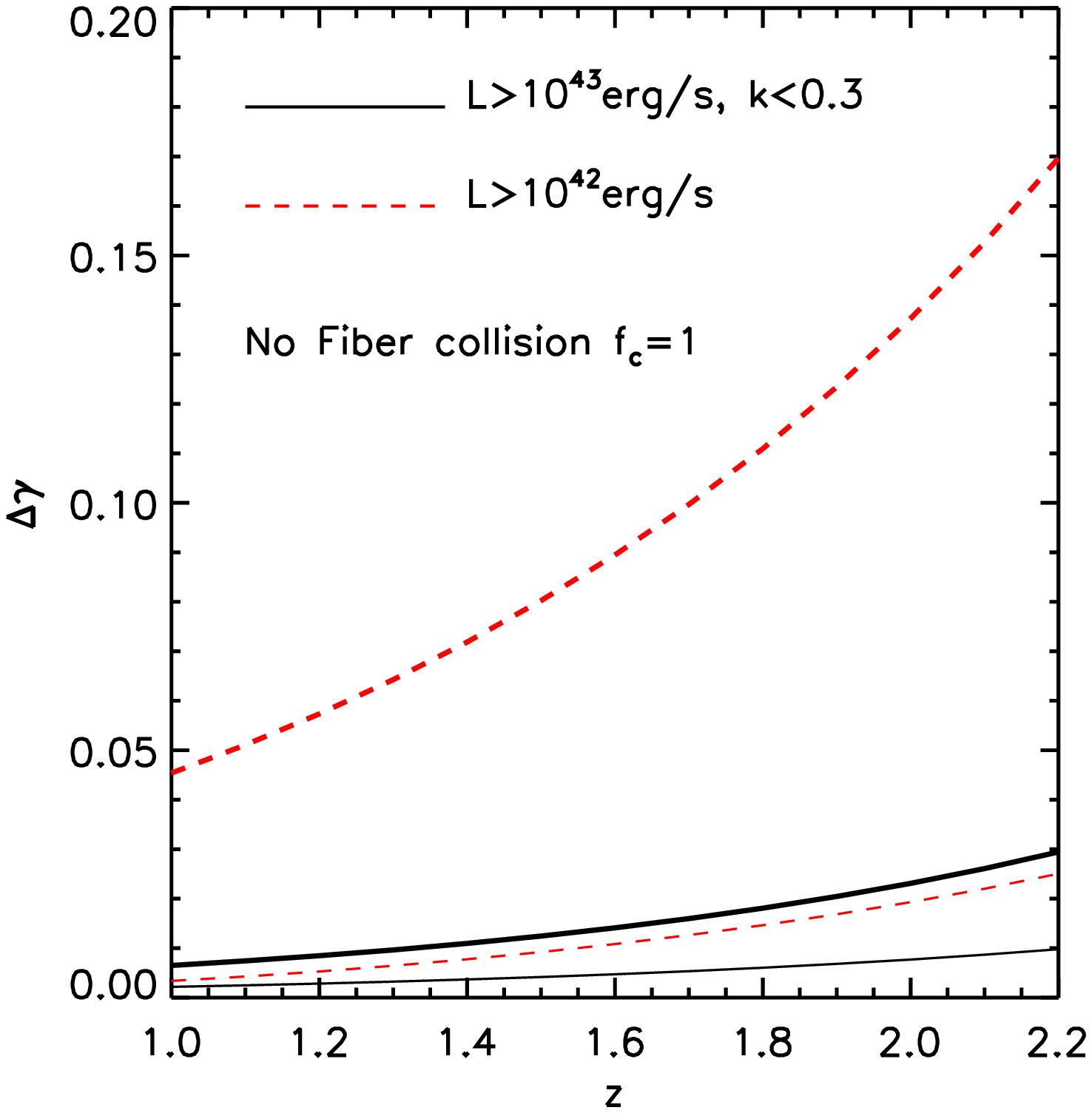}
\caption{$\Delta \gamma$ as a function of the redshift for the
  H$\alpha$ emitter sample $L>10^{43}$erg/s (solid curve) and
  $L>10^{42}$erg/s (dashed curve). The left panel is no fiber
  collision $f_{\rm col}=1$, while the right panel is the case with
  the fiber collision $f_{\rm col}=0.5$.  In each panel, the thin
  curve is the case A estimation, but the thick curve is the case B
  estimation.
\label{Fig:Fisher}
}
\end{center}
\end{figure*}

\subsection{Fisher matrix}
We here discuss about systematic errors from uncertainties of the satellite galaxies 
in future redshift survey at a quantitative level. 
To this end, we adopt the Fisher matrix technique to estimate the systematic errors 
from the one halo term (see, e.g., \cite{Hikage2011,Tayloretal,Knox}). 
The bias in a parameter is estimated by 
\begin{eqnarray}
  \delta\theta_i=-[F^{\theta\theta}]^{-1}_{ik}F_{kj}^{\theta\psi}\delta \psi_j,
\end{eqnarray}
where $F^{\theta\theta}_{ij}$ is the Fisher matrix, whose inverse matrix is 
$[F^{\theta\theta}]^{-1}_{ik}$, and $F_{kj}^{\theta\psi}\delta \psi_j$
is a vector which describes the systematic bias caused by ignoring 
the one-halo term. In {\it case A}, we adopt the expressions 
\begin{eqnarray}
  &&F^{\theta\theta}_{ij}={1\over 8\pi^2}\int_{k_{\rm min}}^{k_{\rm max}} dk k^2
  \int_{-1}^{+1} d\mu {\partial P(k,\mu)\over \partial\theta_i}
  {\partial P(k,\mu)\over \partial\theta_j}
  {V\over [P(k,\mu)+1/\bar n]^2},
\\
  &&F_{ij}^{\theta\psi}\delta \psi_j={1\over 8\pi^2}\int_{k_{\rm min}}^{k_{\rm max}} dk k^2
  \int_{-1}^{+1} d\mu {\partial P(k,\mu)\over \partial\theta_i}{P^{1h}(k,\mu)}
  {V\over [P(k,\mu)+1/\bar n]^2}, 
\end{eqnarray}
where $V$ is a survey volume, and 
we set $k_{\rm min}=0.01h{\rm Mpc}^{-1}$ and $k_{\rm max}=0.3h{\rm Mpc}^{-1}$.
In {\it case B}, we use \cite{YBN}
\begin{eqnarray}
  &&F^{\theta\theta}_{ij}=\sum_{\ell=0,2,\cdots}^{\ell_{\rm max}=6}{1\over 4\pi^2}
\int_{k_{\rm min}}^{k_{\rm max}}
dkk^2 {\partial P_\ell(k)\over \partial\theta_i}{\partial P_{\ell}(k)\over \partial\theta_j}
    \kappa(k),
\\
  &&F_{kj}^{\theta\psi}\delta \psi_j=\sum_{\ell=0,2,\cdots}^{\ell_{\rm max}=6}{1\over 4\pi^2}
 \int_{k_{\rm min}}^{k_{\rm max}}dkk^2 
{\partial P_\ell(k)\over \partial\theta_i}{P^{1h}_{\ell}(k)}
 \kappa(k) 
\end{eqnarray}
with
\begin{eqnarray}
  &&\kappa(k)={1\over 2\ell+1}{V\over [P_0(k)+1/{\bar n}]^2}.
\end{eqnarray}
In the above expressions, we consider the power spectrum that is the
combination of the one-halo term (\ref{eq:pk_1h}) and the two-halo
term (\ref{eq:pk_2h}), with the growth rate $f=\Omega_m(z)^\gamma$,
\begin{eqnarray}
&&b(M)=(b_0+b_1k)b_{\rm halo}(M),
\\
&&\tilde p_{cs}(k,\mu;M)=e^{-\alpha^2\sigma_{v,off}^2(M)k^2\mu^2/2a^2H^2},
\label{tildepcs}
\end{eqnarray}
where $\gamma$, $b_0$, $b_1$, $\alpha$ are parameters, 
and $b_{\rm halo}(M)$ is the halo bias, fixed as \cite{Tinker2010}
\begin{eqnarray}
&&b_{\rm halo}(M)=1-{\nu^a\over \nu^a+\delta_c^a}+0.183\nu^b+0.265\nu^c,
\end{eqnarray}
with $\nu=\delta_c/\sigma(M,z)$, $\delta_c=1.686$, $a=0.132$, $b=1.5$ and $c=2.4$.
We adopt the 4 parameters, $\gamma$, $b_0$, $b_1$, $\alpha_{}$ 
for the Fisher matrix analysis, where the target parameter is 
$\gamma=0.55$, $b_0=1$, $b_1=0.2$, $\alpha_{}=1$.
The background cosmology is fixed to be the $\Lambda$CDM  
model with $\Omega_m=0.3$, $\Omega_b=0.044$, and $\sigma_8=0.8$. 

In the present paper, we focus on the systematic bias in $\gamma$,
which is considered to be useful for testing gravity. Figure
\ref{Fig:Fisher} shows the systematic bias $\Delta \gamma$ as a
function of the redshift.  In each panel, the solid curve (dashed
curve) adopts the HOD with $L>10^{43}$ erg/s ($L>10^{42}$ erg/s), and
the thin (thick) curve is the {\it case A} ({\it case B}) for the
estimation of the Fisher matrix, respectively.  The left panel assumes
no Fiber collision, while the right panel take the fiber collision
into account by assuming $f_{\rm col}=0.5$.

Figure \ref{Fig:Fisher} means that the fiber collision reduces the
systematic bias because the satellite fraction, which causes the
systematic error, is reduced.  Furthermore, brighter H$\alpha$
emitters do not generally contain satellite, which also reduces the
systematic bias. The mean number density is $\bar n\simeq (2\sim
3)\times 10^{-4} (h/{\rm Mpc})^3$ for the H$\alpha$ emitter with
$L>10^{43}$erg/s (solid curve), while $\bar n\simeq (2\sim 3)\times
10^{-3} (h/{\rm Mpc})^3$ for H$\alpha$ emitter with $L>10^{42}$erg/s
(dashed curve).  The number density of galaxies of a optimized
redshift survey would be $\bar n\simeq (2\sim 3)\times 10^{-4} (h/{\rm
  Mpc})^3$.  In this case, the sample with $L>10^{43}$erg/s (solid
curve) will be a realistic sample, whose systematic bias in $\gamma$
is not large.  It might be worthy to note that an analysis with the
multipole power spectrum ({\it case B}: thick curve) makes a larger
systematic bias compared with an analysis with the full anisotropic
power spectrum ({\it case A}: thin curve).

In general, the amplitude of the one-halo term becomes smaller at
higher redshift because the halo mass becomes smaller.  However, the
power spectrum is less sensitive to the cosmological parameter at
higher redshift, which reduces the Fisher matrix elements at higher
redshift. This is one of the reason why the systematic bias becomes
larger at higher redshift.  In the present paper, we omitted the
random velocity dispersion between halos.  For the H$\alpha$ emitters,
however, the halo random velocity could be large.  This effect will be
included in the two halo term, but not in the one halo term.  Then,
this might not be included as an uncertainty of the one halo term, but
is related with the modeling of the two halo term.  A more precise
theoretical model for the H$\alpha$ emitters will be necessary
including the HOD model and the fiber collision, depending on
observational strategy. Our results here are obtained by extensively
using the HOD model, which was originally obtained at $z>2$.

\section{Summary and Conclusions} 
\label{sec:conclusion}
In the present paper, we have investigated the influence of the
satellite galaxies on the redshift-space distortions.  We have found
the following points, for the first time.  First, the satellite
galaxies significantly contribute to the higher-order multipole power
spectrum though the fraction is small.  Second, the contribution of
the satellite galaxies to the higher-order multipole power spectrum is
explained by a simple halo model, and the one halo term makes the
dominant contribution.  We have also demonstrated that the
contribution from satellite galaxies depends on the HOD of galaxy
samples and the effect of the fiber collision.  These findings are
based on the SDSS LRG sample, but generally means that an uncertainty
of the HOD might give rise to a systematic error in measuring
redshift-space distortion when satellite galaxies are contaminated.
We have also demonstrated that the small-scale information of higher
multipole spectra $P_4(k)$ and $P_6(k)$ at large wavenumbers help
calibrate the satellite FoG effect and improve the measurement of
growth rate dramatically.

For the H$\alpha$ emitters, which are the target galaxies of the PFS
redshift survey and the Euclid redshift survey, we have shown that the
satellite's contribution to the redshift-space distortion is much
smaller than the case of LRGs, because the host halo mass is small.
The results are based on the HOD of the H$\alpha$ emitters at the
redshift $z>2$, it would be interesting to investigate how the results
change depending on the redshift especially in the lower redshift
regions.  Combination with weak lensing survey might help to resolve
the uncertainty in HOD \cite{FCB1,FCB2,FCB3}.  A simple Fisher matrix
analysis shows that the systematic error from the HOD uncertainty in
the parameter $\gamma$ is not large for H$\alpha$ emitters with
$L>10^{43}$ erg/s. But this conclusion is based on the simple model
with the HOD model, which was originally obtained at $z>2$.  Then
further check will be necessary, including a modeling of the peculiar
velocity of halos.

The one-halo term makes the significant contribution to the higher
multipole power spectrum of the LRG sample.  It is expected that the
same situation happens in the CMASS sample of the BOSS survey.  The
one-halo term reflects the HOD as well as the random velocities of
satellite galaxies in a halo.  This fact might provide us with an
additional cosmological information on the scales of cluster of
galaxies.  For example, in a class of modified gravity model, the
effective gravitational constant in a halo could be larger than that
of the solar system. This enhances the velocity of satellite galaxies,
which might be detected a signature of modified gravity theories
(c.f. \cite{Nishimichi2012}).  Such a signature might be constrained
from the observation of higher multipole power spectrum. But we have
also demonstrated that such a gravity-test requires the precise
information of the velocity probability distribution function of
satellite galaxies as well as the HOD, plus the fiber collision
effect.  This subject is also left as a future problem.

\acknowledgments We thank M. Takada and S. Masaki for useful
discussions at the early stage of this work.  We also thank A. Oka,
S. Saito, T. Nishimichi, A. Taruya, T. Matsubara, T. Okumura,
T. Kanemaru, and A. Terukina for useful communications related to the
topic of the present paper. We acknowledge anonymous referree for
useful and constructive comments. The research by K.Y. and C.H. is
supported in part by Grant-in-Aid for Scientific researcher of
Japanese Ministry of Education, Culture, Sports, Science and
Technology (No.~21540270 and No.~21244033 for K.Y. and No.~24740160
for C.H.). K.Y. is also supported by exchange visitor program between
JSPS and DFG.

\appendix
\section{Derivation of power spectrum}
In this appendix, we derive a general expression of the multipole
power spectrum in the halo model, which gives the grounds to adopt the
expressions in section 3.  Following the halo model approach, the
correlation function is written as the sum of the 1-halo term and the
2-halo term.  The power spectrum is the Fourier transform of the
correlation function, then the power spectrum is also written as the
combination of the 1-halo term and the 2-halo term.  We start with the
real-space power spectrum in a halo model presented in
reference~\cite{Skibba},
\begin{eqnarray}
P^{R}(k)=P^{R1h}(k)+P^{R2h}(k),
\end{eqnarray}
where we defined
\begin{eqnarray}
&&P^{R1h}(k)={1\over \bar n^2}\int dM{dn(M)\over dM}\langle N_{cen}\rangle
\biggl[2\langle N_{sat}\rangle \tilde u_{\rm NFW}(k;M)+\langle N_{sat}(N_{sat}-1)\rangle 
\tilde u_{\rm NFW}(k;M)^2\biggr],
\nonumber
\\
\\
&&P^{R2h}(k)={1\over \bar n^2}
\left[\int dM{dn(M)\over dM}\langle N_{cen}\rangle
\left(1+\langle N_{sat}\rangle \tilde u_{\rm NFW}(k;M)\right)b(M)\right]^2P_{m}(k),
\end{eqnarray}
and $\tilde u_{\rm NFW}(k;M)$ is the Fourier transform of the density profile of galaxy
distribution.  We assume that the galaxy density profile is the same
as the dark matter density profile.  For the NFW density profile, we
have \cite{Scoccimarro2001,CooraySheth2002}
\begin{eqnarray}
\tilde u_{\rm NFW}(k;M)&=&{\int_{r\leq r_{vir}}d^3x \rho(x|M) e^{-i{\bf k}\cdot{\bf x}}\over
\int_{r\leq r_{vir}}d^3x \rho(x|M)}
\nonumber
\\
&=&{4\pi \rho_s r_s^3\over M}
\biggl\{ \sin (kr_s) \left[ Si([1+c]kr_s)-Si(kr_s)\right] \nonumber \\
&&-{\sin ckr_s\over(1+c)kr_s}
+\cos (kr_s) \left[ Ci([1+c]kr_s)-Ci(kr_s)\right]\biggr\}, 
\label{tuNFW}
\end{eqnarray}
where 
\begin{eqnarray}
C_i(x)=-\int_x^\infty {\cos t\over t} dt,~~~~~~
S_i(x)=\int_0^x {\sin t\over t} dt.~~~~~~
\end{eqnarray}
The redshift-space power spectrum of the halo model may be evaluated
as follows.  Tinker investigated the formulation for the
redshift-space correlation function in a halo model \cite{Tinker2}, in
which the redshift-space correlation function is obtained by
\cite{Scoccimarro2004,Peebles1979} 
\begin{eqnarray}
\xi(s_\perp,s_\parallel)=\int\xi^{R}\left(r\right) {\cal P}(v_z)dv_z,
\end{eqnarray}
where $\xi^R(r)$ is the real-space correlation function, $s_\perp$ is
the projected separation, $s_\parallel$ is the line of sight
separation, $r^2=s^2_\perp+z^2$ and $v_z=H(s_\parallel-z)$, ${\cal
  P}(v_z)$ is the probability distribution function of the galaxy
pairwise velocity, and $H$ is the Hubble parameter.  ${\cal P}(v_z)$
maps the pairs at separation in the line-of-sight direction $z$ to
$s_{\parallel}$ with the probability ${\cal P}(v_z)$
\cite{Scoccimarro2004,Peebles1979}.  This gives the prescription to
include the random velocity of galaxies in a halo in redshift-space
power spectrum.  Then, we may write the redshift-space power spectrum
in the form
\begin{eqnarray}
P(k,\mu)=P^{1h}(k,\mu)+P^{2h}(k,\mu),
\end{eqnarray}
where 
\begin{eqnarray}
P^{1h}(k,\mu)={1\over \bar n^2}\int dM{dn(M)\over dM}\langle N_{cen}\rangle
\left[2\langle N_{sat}\rangle \tilde p_{cs}(k,\mu;M)
\right.
~~~~~~~ \nonumber \\
\times \left.
+\langle N_{sat}(N_{sat}-1)\rangle \tilde p_{ss}(k,\mu;M)\right],
\label{ap1ht}
\end{eqnarray}
and
\begin{eqnarray}
P^{2h}(k,\mu)=
\biggl[{1\over \bar n}
\int dM{dn(M)\over dM}\langle N_{cen}\rangle ~~~~~~~~~~~~~~~~~~~~~~~~~~~~~~~~~~~
\nonumber\\
\times\left(1+\langle N_{sat}\rangle \tilde p_{cs}(k,\mu;M)\right)
(b(M)+f\mu^2)\biggr]^2P_{m}(k), 
\label{ap2ht}
\end{eqnarray}
where we defined
\begin{eqnarray}
\tilde p_{cs}(k,\mu;M)=\tilde u_{\rm NFW}(k;M)e^{-\sigma_{v}^2k^2\mu^2/2a^2H^2}, ~~~~~
\end{eqnarray}
when the pair wise velocity between the central galaxy and the satellite galaxy 
obeys the Gaussian probability distribution function
${\cal  P}(v_z)=(\sqrt{2\pi}\sigma_v)^{-1} e^{-v_z^2/2\sigma_v^2}$. 
Here we assume that the random velocity of the central galaxies can be 
neglected, then we may write $\tilde p_{ss}(k,\mu;M)=\tilde p_{cs}^2(k,\mu;M)$
for the satellite-satellite galaxy pair. 
In the case of the exponential velocity distribution function,
${\cal  P}(v_z)=(\sqrt{2}\sigma_v)^{-1} e^{-\sqrt{2}|v_z|/\sigma}$, we have
\begin{eqnarray}
&&\tilde p_{cs}(k,\mu;M)={\tilde u_{\rm NFW}(k;M)\over 1+\sigma_{v}^2k^2\mu^2/2a^2H^2}
=\tilde u_{\rm NFW}(k;M){\cal D}\Bigl({\sigma_{v}k \mu\over aH}\Bigr), 
\label{pscexpv}
\\
&&\tilde p_{ss}(k,\mu;M)={\tilde u_{\rm NFW}(k;M)^2\over 1+\sigma_{v}^2k^2\mu^2/a^2H^2}
=\tilde u_{\rm NFW}(k;M)^2{\cal D}\Bigl({\sqrt{2}\sigma_{v}k \mu\over aH}\Bigr).
\label{pscexpv2}
\end{eqnarray}

As is shown in section 3, the one-halo term dominates the higher multipole power 
spectrum of the All LRG sample. It is useful to present the analytic formula, as
is given by equation~(\ref{P1hlexpress}) with (\ref{Q0express})-(\ref{Q6express}) for
the case of the Gaussian velocity distribution function. 
In the case of the exponential velocity distribution function, 
(\ref{Q0express})-(\ref{Q6express}) are replaced with 
\begin{eqnarray}
Q_0(q)&=&\frac{\displaystyle \arctan q}{\displaystyle q}, 
\label{Q0expresss}
\\
Q_2(q)&=&\frac{\displaystyle 3q-(3+q^2)\arctan q}{\displaystyle 2q^3}, 
\\
Q_4(q)&=&\frac{\displaystyle -105 q-55 q^3+(105+90q^2+9q^4)\arctan q}{\displaystyle 24q^5}, 
\\
Q_6(q)&=&\frac{\displaystyle 1155q+1190 q^3+231q^5-(1155+1575q^2+525q^4+25q^6)\arctan q}
{\displaystyle 80q^7}.
\label{Q6expresss}
\end{eqnarray}


\end{document}